\documentclass[pra, reprint, 10pt,british]{revtex4-1}
\usepackage[T1]{fontenc}
\usepackage[latin9]{inputenc}
\usepackage[letterpaper]{geometry}
\usepackage{mathrsfs}	

\usepackage{color}
\usepackage{amsbsy}
\usepackage{amssymb}
\usepackage{graphicx}
\usepackage{esint}

\makeatletter

\providecommand{\tabularnewline}{\\}

\usepackage{subfig}
\makeatother

\usepackage{babel}
\begin{document}


\title{General phenomenology of ionization from aligned molecular ensembles}

\author{Paul Hockett}
\email[Email: ]{paul.hockett@nrc.ca}
\affiliation{National Research Council of Canada, 100 Sussex Drive, Ottawa,K1A 0R6, Canada}

\date{05/09/14}

\begin{abstract}
Single and multi-photon ionization of aligned molecular ensembles
is examined, with a particular focus on the link between the molecular
axis distribution and observable in various angle-integrated and angle-resolved
measurements. To maintain generality the problem is treated geometrically,
with the aligned ensemble cast in terms of axis distribution moments,
and the response of observables in terms of couplings to these moments.
Within this formalism the angular momentum coupling is treated analytically,
allowing for general characteristics - independent of the details
of the ionization dynamics of a specific molecule - to be determined.
Limiting cases are explored in order to provide a phenomenology which
should be readily applicable to a range of experimental measurements,
and illustrate how observables can be sensitive to fine details of
the alignment, i.e. higher-order moments of the axis distribution,
which are often neglected in experimental studies. We hope that this
detailed and comprehensive treatment will bridge the gap between existing
theoretical and experimental works, and provide both quantitative
physical insights and a useful general phenomenology for researchers
working with aligned molecular ensembles.

\end{abstract}
\maketitle

\section{Introduction}

Over the last decade techniques for molecular alignment have become
increasingly advanced, and increasingly popular. Perhaps the most
common, and experimentally accessible, technique is non-adiabatic
(or impulsive) alignment, achieved via the interaction of a short, intense
laser pulse with a rotationally cold, gas phase sample \cite{Seideman1995,Rosca-Pruna2001,Renard2003,Stapelfeldt2003,Seideman2006}.
In the case of an IR driving field, multiple cascaded Raman transitions
occur during the pulse, populating many rotational levels, thereby
creating a broad rotational wavepacket in the system. After the pulse,
the wavepacket propagates under field-free conditions, and undergoes
revivals at characteristic times, determined by the rotational constants
of the molecule. The utility of this type of alignment - as compared
to adiabatic alignment techniques - is that further experiments may
be carried out in the vicinity of the revivals, thus providing field-free
conditions for these measurements, albeit on a highly rotationally
excited system.

Recent examples of the application of this technique span a wide gamut
of measurements, including weak and strong-field photoelectron angular
distributions for probing dynamics \cite{Reid1999,Tsubouchi2001,Suzuki2006}
or ``approaching the molecular frame'' \cite{Kumarappan2008,Bisgaard2009,Holmegaard2010,Maurer2012,Rouzee2012},
high-harmonic measurements \cite{Vozzi2011,Lock2012,Lin2012,Bertrand2012,Frumker2012,Ren2013},
angle-resolved ATI \cite{Pavicic2007,Mikosch2013a}, coulomb explosion
\cite{Dooley2003,Lee2006a,Rouzee2012} and X-ray diffraction \cite{Kupper2013}
to mention just a few examples. In most cases the application of
alignment is at the qualitative level, where the alignment is optimized
based on a proxy for the degree of alignment (e.g. ionization yield
at a revival feature of the rotational wavepacket \cite{Mikosch2013a,Mikosch2013}),
and the experimental goal is to maximize the alignment effect, or
observe some phenomena which would otherwise be obscured by orientational
averaging - for instance imaging torsional motions \cite{Hansen2012}.
In other cases, the aim is a more quantitative study of the rotational
wavepacket behaviour \cite{Mikosch2013,Ramakrishna2013}, or detailed
understanding of molecular frame phenomena which can be directly observed
if the degree of alignment is high \cite{Kupper2013}, or may be ``extracted''
from measurements in the lab frame in certain cases provided the alignment
is well-characterized and the coupling to the observable well-understood
\cite{Underwood2000,Bertrand2012,Suzuki2012,Ramakrishna2012}.

In terms of a qualitative approach, the degree of alignment is often
considered solely in terms of $\langle\cos^{2}\theta\rangle$ (the expectation value of $\cos^{2}\theta$), where $\theta$ is the angle between the molecular axis and lab frame $z$-axis \cite{Stapelfeldt2003},
and this metric is treated
as completely defining the axis distribution in the lab frame.
For a fuller treatment of the molecular axis distribution, higher-order
moments of the distribution (e.g. $\langle\cos^{n}\theta\rangle$) must be taken into account. For example, photoelectron angular distributions (PADs) are known to be sensitive
to higher-order alignment moments \cite{Underwood2000,Suzuki2005}.
This response of higher-order observables to higher-order alignment
moments has long been implicit in work on photoionization, and was
first discussed explicitly in the context of recent work on rotational
wavepackets by Seideman \cite{Seideman1997} and Underwood \& Reid
\cite{Underwood2000}. Some of the implications of this coupling have
been investigated extensively in theory work from Seideman \& co-workers
\cite{Seideman2000,Suzuki2005,Ramakrishna2012,Ramakrishna2013}, the
most recent of which discusses the possibility of mapping alignment
via measurement of photoelectron angular distributions as well as
the use of other probe techniques.

Regardless of the aims of a given study, but of particular importance
in experiments seeking molecular frame properties, is the detailed
understanding of the probe process, and the response of the observable
to the degree of alignment. At a basic level this is required to formulate
metrics to optimise the degree of alignment and interpret results
in terms of the underlying properties of interest. In the lowest-order
approach, the optimum alignment corresponds to maximizing $\langle\cos^{2}\theta\rangle$
 and, practically, such optimization typically takes the form of
maximising the contrast observed between the alignment and anti-alignment
features of a revival of the rotational wavepacket. However, depending
on the observable, the contrast may be poor or negligible, even in
the presence of a highly aligned distribution, negating the use of
such signals as a measure of alignment \footnote{The illustration of this effect is one of the purposes of the current work, but we note in passing that this issue is discussed in terms of photoelectron yields in ref. \cite{Reid2000}, wherein a certain choice of ionization parameters was shown to result in effectively no response of the yield to molecular axis alignment, and in terms of ion fragment yields in ref. \cite{Rosca-Pruna2001}, wherein different fragment channels were found to exhibit markedly different revival contrast. In this latter case, the observation was ascribed to the ionization channel dependence on probe laser intensity and concomitant focal-volume averaging effects.}. In these cases a more detailed
treatment of the alignment may be necessary for even a qualitative
interpretation of experimental results.

At a more detailed level one might hope to fully characterize the
aligned distribution, allowing for more quantitative analysis of experimental
data. This is a non-trivial task but, despite the complications,
such detailed analysis has been attempted in a few cases, for example
refs. \cite{Suzuki2006,Pavicic2007,Vozzi2011,Rouzee2012,Suzuki2012,Lock2012,Mikosch2013}
all illustrate detailed analyses of the prepared alignment and various
molecular properties. In tandem with the experimental efforts, various
\emph{ab initio} studies have also presented results for specific
molecules and types of measurements (see, for example, refs. \cite{Arasaki2001,Le2009,Abdurrouf2009,Jin2010,Ramakrishna2013}).
It is interesting to note that in many cases (e.g. single photon ionization,
fluorescence) the fundamentals have long been known, but have only
been applied in older studies where static alignments, or narrow rotational
wavepackets, were prepared.
\footnote{We note for completeness that there is a strong relation of much recent
work to rotational coherence spectroscopy (RCS), as developed by Felker
\& Zewail \cite{Felker1987}. The primary difference between older
and more modern alignment techniques is the preparation of the rotational
wavepacket using strong IR pulses, leading to much broader rotational
wavepackets than those created via single (or few) photon absorption.
Additionally, the low-order observables in RCS render the observed
signals sensitive to only low-order moments of the aligned distribution.
Similar comments hold when comparing frequency-resolved photoionization
measurements to time-resolved measurements: the underlying physics
in terms of rotational couplings is identical, but the narrow rotational
distributions prepared are phenomenologically quite different.%
}

In this work we also investigate the response of observables in photoionization
experiments to aligned distributions, and approach the problem quite
generally from a geometric perspective. A geometric approach allows
for the separation of the molecular axis distribution from other,
molecule specific, properties \cite{Dill1976,Underwood2000,Stolow2008}.
We thus aim to provide a useful and general applied phenomenology
which can provide qualitative and quantitative insights into the ionization
of aligned distributions. This treatment begins with the formalism
of Underwood \& Reid \cite{Underwood2000} (which was further discussed
and extended in Stolow \& Underwood \cite{Stolow2008}), from this
we derive and discuss explicit forms for typical experimental measurements.
We further extend the formalism to $N$-photon ionization, in order
to discuss the link between single and multi-photon ionization processes,
including both angle-integrated and angle-resolved observables. Typically,
single and multi-order processes are treated independently by the
``weak'' and ``strong'' field communities despite the similarities
of the underlying physical processes, so such a treatment may be useful
in bridging this divide in some cases.%
\footnote{It is the case, however, that the treatment developed here is strictly
valid only in the perturbative regime, so can only be assumed to be
qualitative in the true strong-field regime. For multi-photon processes
at moderate intensities, however, it should be applicable.%
} Using the formalism presented we explore the general form of different
experimental measurements - specifically angle-integrated ionization
yields as a function of pump-probe polarization geometry, polarization-angle
resolved measurements, and angle-resolved photoelectron measurements
- and, by incorporating rotational wavepacket calculations for an
example system under typical experimental conditions, investigate
limiting cases, providing an aid to experimentalists working to prepare
and optimise aligned distributions. 

Although some aspects of this work are extant in the literature and
known by practitioners in either the alignment community or the photoionization
community, not to mention other fields which make use of strong-field
alignment (as noted above), the complexity of the wavepacket and observables
have resulted in very few works which cross-over between these fields
and cover all aspects from wavepacket to observable. Notable exceptions
are experiments from Suzuki and co-workers \cite{Tsubouchi2001,Suzuki2006,Tang2010},
and theory from Underwood \& Reid \cite{Underwood2000}, and Seideman
and co-workers \cite{Seideman1997,Seideman2002,Ramakrishna2012,Ramakrishna2013}.
In a following paper \cite{hockett2014b} we apply our formalism
to ionization of aligned butadiene and compare theory with experimental
results; hence, we aim here to provide complementary insights and
more direct comparison to experiment than existing studies, as well
as a unified and generalised geometric treatment of the relevant theory. Most generally, we aim to bridge
the gap between theoretical understanding of the physics, notably
the angular momentum coupling between the axis distribution and the observables,
and experiments where one must treat specific cases, usually with
many unknowns, in particular the precise, molecule-specific details of the probe process.

\section{Theory}

The necessary theory for a full treatment of molecular alignment and
ionization is discussed here, with a focus on quantification of the
relevant parameters. We begin by treating the aligned distribution
geometrically for the most general case (full 3D axis alignment),
and the simpler cases of 2D or 1D distributions which are applicable
to symmetric top \& linear molecules, and experimental configurations
with cylindrical symmetry. We then consider parametrization of alignment
in detail, and the coupling of these alignment metrics into ionization
measurements, for single and multi-photon ionization processes,
including frame rotations and the response of the observables in typical
experimental measurements. Throughout example calculations are used
to illustrate the equations, and investigate limiting cases, which are further discussed in sect. \ref{sec:Application}.

\subsection{Axis distribution moments\label{sub:Axis-distribution-moments}}

We begin with a description of the spatial distribution of molecular
axes. Most generally, the distribution should be described by an expansion
in 3D functions, specifically the Wigner rotation matrices \cite{Stolow2008}:
\begin{equation}
P(\Omega,t)=\sum_{K,Q,S}A{}_{Q,S}^{K}(t)D_{Q,S}^{K}(\Omega)\label{eq:Ptheta_D}
\end{equation}
where $\Omega=\{\Phi,\Theta,\chi\}$ are the Euler angles describing the position
of the molecular axis relative to a reference frame, $D_{Q,S}^{K}(\Omega)$
are Wigner rotation matrices and $A{}_{Q,S}^{K}(t)$ are expansion
parameters which are generally termed the axis distribution moments
(ADMs) and, for dynamical systems, may be time-dependent. The quantum
numbers $K,\, Q,\, S$ denote the rank/moment of the distribution,
and projections onto the space and body-fixed axes respectively.

The general form of the distribution can be simplified for the case
of linear and symmetric top molecules, for which $S=0$ only, and
the Wigner matrices can be replaced with an expansion in 2D functions:

\begin{equation}
P(\theta,\phi,t)=\sum_{K,Q}A_{K,Q}(t)Y_{K,Q}(\theta,\phi)\label{eq:Ptheta_YKQ}
\end{equation}
where $Y_{K,Q}(\theta,\phi)$ are spherical harmonics. (Throughout
this manuscript we use upper-case $\Theta,\,\Phi$ for Euler angles,
and lower-case $\theta,\,\phi$ for spherical polar coordinates although,
in many cases, the angles are referenced to the same axis, hence are
identical.)

In the case of a cylindrically symmetric distribution, for which $Q=0$,
a further simplification to use 1D functions can be made:

\begin{equation}
P(\theta,t)=\sum_{K}A_{K}(t)\mathscr{P}_{K}(\cos(\theta))\label{eq:Ptheta_PL}
\end{equation}
where $\mathscr{P}_{K}(\cos(\theta))$ are Legendre polynomials in $\cos(\theta)$, and normalization factors relative to the spherical harmonic expansion above are subsumed into the $A_{K}$.

In practice, the laboratory frame of reference (LF) polar axis ($z$)
is chosen to be defined by the laser polarization axis for linearly
polarized light, or the propagation axis for elliptically (or pure
circularly) polarized light, of the (pump) laser pulse used to prepare
the aligned sample. In the former case the axis distribution is constrained
to be cylindrically symmetric, and in both cases there is reflection
symmetry along the polar ($z$) axis, so $K=0,2,4,6....K_{max}$ (odd
$K$ can only appear in the case of an oriented distribution). 

In the remainder of this work we restrict our discussion to the case
of linearly polarized laser pulses for simplicity, although the
formalism given here can be applied to any arbitrary polarization
state. For linearly polarized light the axis distributions are defined
by the cylindrically symmetric $P(\theta,t)$, as given by eqn. \ref{eq:Ptheta_PL},
although the symmetry may be broken in the case of a frame rotation between pump and probe
pulses (see below), which can lead to a $\phi$ dependence of the
axis distribution in the probe reference frame. To allow for this, we use the spherical harmonic expansion given by eqn. \ref{eq:Ptheta_YKQ} as the more general definition throughout this work, even when discussing cylindrically symmetric distributions ($Q=0$, no $\phi$ dependence), but omit the $\phi$ label in cylindrically symmetric cases.

\begin{figure}

\includegraphics{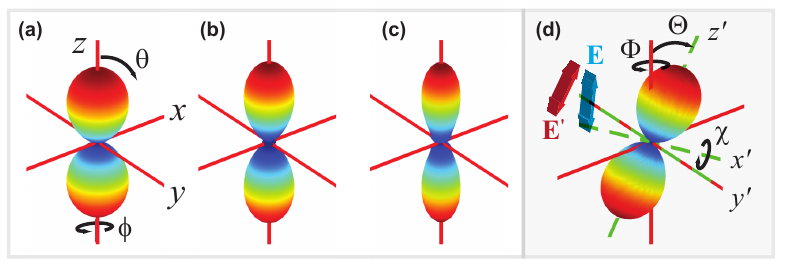}

\caption{Example axis distributions $P(\theta,\phi)$ for distributions with
(a) $A_{2,0}=1$, (b) $A_{2,0}=1,\, A_{4,0}=0.4$ and (c) $A_{2,0}=1,\, A_{4,0}=0.4,\, A_{6,0}=0.3$.
(d) shows the distribution in (a) after a frame rotation of $\Theta$
between the pump (alignment) pulse and the probe (ionization) pulse, with polarization vectors $\mathbf{E'}$ and $\mathbf{E}$ respectively;
the polarization geometry of the measurement is thus defined by $\Theta$.
The spherical polar coordinate system $(\theta,\phi)$ and the Euler
angles $\Omega=\{\Phi,\Theta,\chi\}$ are also shown in (a) and (d)
respectively.\label{fig:Example-axis-distributions}}
\end{figure}

Examples of cylindrically symmetric axis distributions with increasing $K_{max}$ are shown
in figure \ref{fig:Example-axis-distributions}(a) - (c), and the
effect of a frame rotation between pump and probe polarization axes
for linearly polarized pulses is shown in figure \ref{fig:Example-axis-distributions}(d).

\begin{figure}
\includegraphics{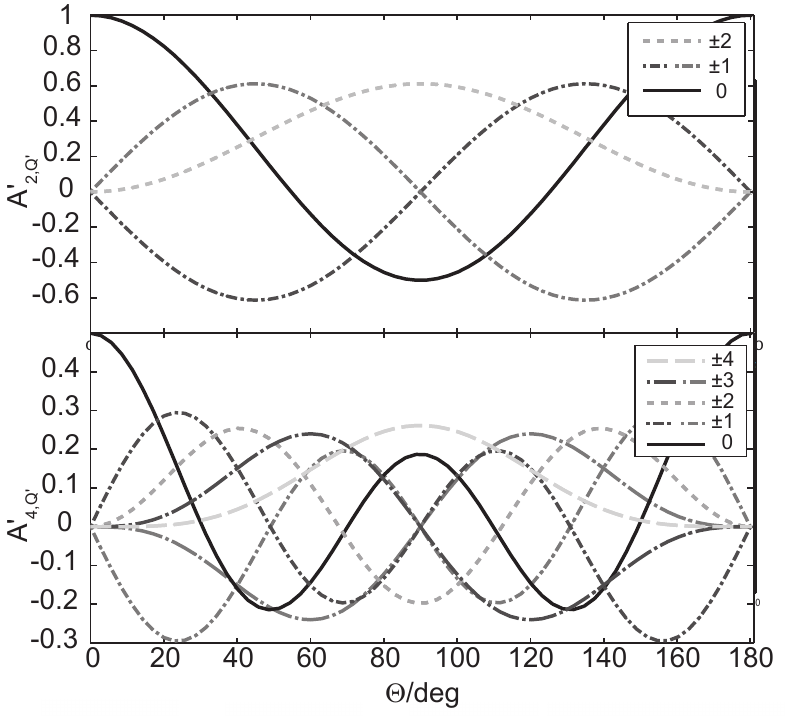}

\caption{Behaviour of axis distribution moments $A'_{K,Q'}$ under frame rotation of $\Theta$
for a distribution with $A_{2,0}=1,\, A_{4,0}=0.5$ in the initial,
unrotated frame ($\Theta=0$). Note that curves for $\pm Q'$ are sign invariant
for even $Q'$.\label{fig:ADMs-under-rotation}}
\end{figure}

The effect of a frame rotation from pump to probe frames, as illustrated
in figure \ref{fig:Example-axis-distributions}(d), can be expressed
in terms of the original and final ADMs and the rotation matrix element
which transforms between the frames according to the set of Euler
angles $\Omega$:

\begin{eqnarray}
A'_{K,Q'}(t;\,\Omega) & = & \sum_{K}\sum_{Q}D_{Q',Q}^{K}(\Omega)A_{K,Q}(t)\\
 & = & \sum_{K}D_{Q',0}^{K}(\Omega)A_{K,0}(t)
\end{eqnarray}
where the second line includes the assumption that $Q=0$, i.e. the
initial distribution is cylindrically symmetric. Even in the case
of such a distribution, the probe frame may contain terms with $Q'\neq0$.
Figure \ref{fig:ADMs-under-rotation} shows the application of a frame
rotation by $\Theta$ on the ADMs illustrated in figure \ref{fig:Example-axis-distributions}(b) ($A_{2,0}=1,\, A_{4,0}=0.5$).
Such rotations are important for consideration of the mapping of rotational
wavepackets, and the understanding of angle-resolved measurements, since
they appear through the dependence of the observable on the polarization
geometry of the measurement.

\subsection{Rotational wavepackets}

In order to calculate the axis distributions defined geometrically
above, for a specific molecular system, knowledge of the rotational
wavefunction is required. In alignment experiments as discussed above,
a rotational wavepacket $\psi(t)$ is prepared via interaction with
an intense IR pump laser pulse, and then evolves under field-free
conditions. 

The axis distribution is determined by the projection of this wavefunction
onto the Euler angles $\hat{\Omega}$, and integrated over the unobserved
angles. For the cylindrically symmetric case (i.e. $\phi$ and $\chi$
are summed over) this is given by \cite{Bisgaard2006}:

\begin{eqnarray}
P(\theta,\, t) & = & \iintop_{0}^{2\pi}|\langle\hat{\Omega}|\psi(t)\rangle|^{2}d\Phi d\chi\label{eq:ptheta}
\end{eqnarray}

The rotational wavefunction can be expanded in the symmetric-top basis:

\begin{equation}
|\psi(t)\rangle=\sum_{J}c_{J}(t)|JKM\rangle\label{eq:psiJKM}
\end{equation}
where $J,\, K$ and $M$ are the usual symmetric-top quantum numbers,
denoting rotational angular momentum $J$ and projections $K$ and
$M$ onto the molecular and laboratory frame $z$-axes respectively;
$t$ is the time, where $t=0$ (henceforth denoted $t_{0}$) is defined
by the peak of the pump laser pulse and the end of the laser pulse
by $t_{f}$; the $c_{J}(t)$ are the expansion coefficients.

Because the $|JKM\rangle$ state populations do not change after the
interaction with the laser pulse, all field-free temporal evolution
of the axis distribution ($t>t_{f}$) is contained in the phase of
the $c_{J}$s, and is given by:

\begin{equation}
c_{J}(t)=c_{J}(t_{f})e^{2\pi E_{J'}t}\label{eq:cjt}
\end{equation}

The calculation of $P(\theta,\, t)$ therefore depends on the calculation
of the $c_{J}(t)$, which requires knowledge of the rotational energy
levels $E_{J}$ and the $c_{J}(t_{f})$. This final aspect, determined
by the light-matter interaction during the pump laser pulse ($t<t_{f}$),
must be treated numerically and is discussed in appendix \ref{sub:Numerics}.

\begin{figure}
\includegraphics{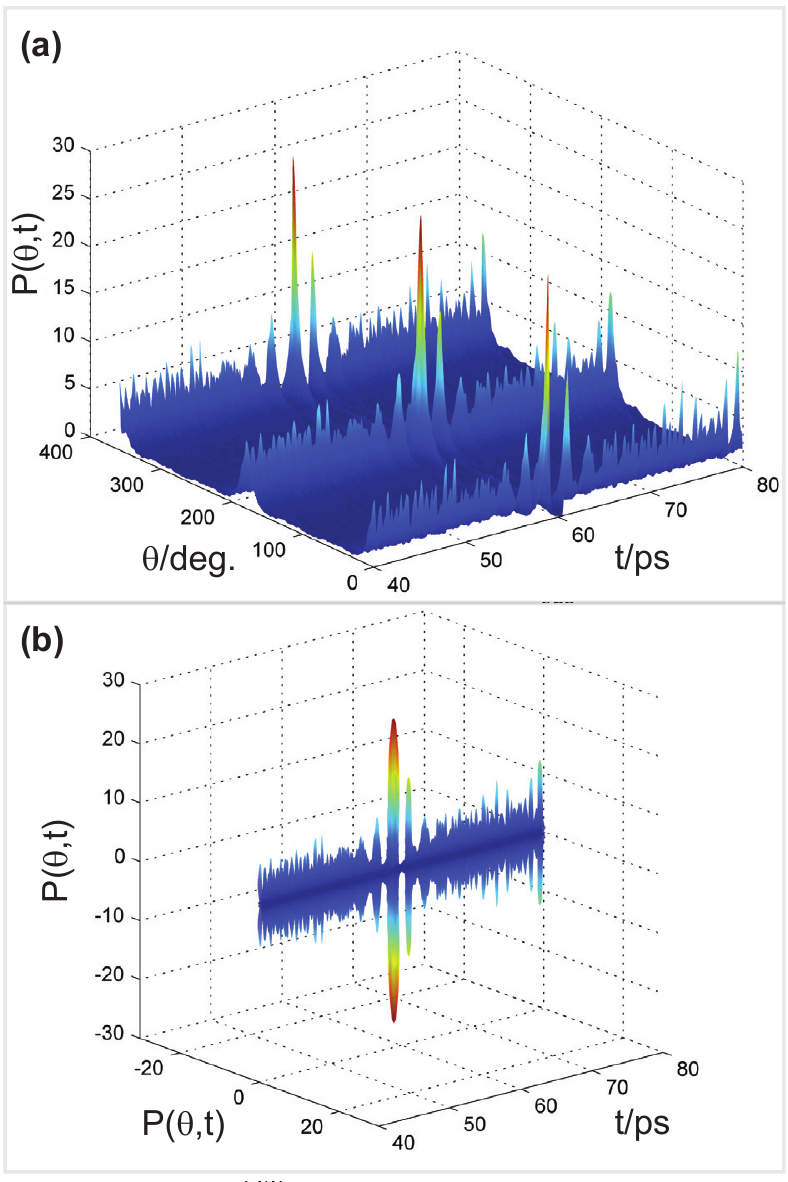}

\caption{Calculated $P(\theta,t)$ for butadiene in the vicinity of the half-revival of the rotational wavepacket, results are shown as (a) Cartesian plot and (b) Polar plot.\label{fig:Ptheta_cart_pol}}

\end{figure}

In this work we use butadiene as a specific example, and base calculations
of $P(\theta,t)$ on ``typical'' experimental conditions (peak intensity (I) = 5
TWcm$^{-2}$, pulse length ($\tau$) = 400~fs, rotational temperature ($T_{r}$) = 2~K), chosen to correspond to recent experimental work on butadiene (which will be discussed
in a later publication \cite{hockett2014b}). In the calculations
we assume a symmetric top molecule, the relevant rotational constants
and polarizabilities for butadiene are given in appendix \ref{sub:Numerics}.
The calculated $P(\theta,t)$ in this case is shown in figure \ref{fig:Ptheta_cart_pol},
in both Cartesian and polar forms \footnote{For clarity, we note that although asymmetric tops are covered by the analytical framework discussed above, due to the symmetric top treatment employed numerically the results illustrated here  will not show any effects associated with asymmetric top rotational wavepacket dynamics \cite{Holmegaard2007}.}. Full discussion of the results, and the coupling of $P(\theta,t)$ into the observables defined in
the remainder of this section, can be found in section \ref{sec:Application};
here we note simply that the distribution exhibits a high degree of
spatial anisotropy, with population heavily weighted to the poles,
indicating high-order $A_{K,Q}$ terms are present (c.f. the low-order
distributions shown in figure \ref{fig:Example-axis-distributions}),
and evolves rapidly along the temporal coordinate.

\subsection{Alignment metrics\label{sec:metrics}}

As discussed above, the degree of alignment is often quantified and
reported in the literature in terms of $\langle\cos^{2}(\theta,t)\rangle$,
the expectation value of $\cos^{2}(\theta)$ at time $t$. This metric
of the axis alignment is defined by \cite{Bisgaard2006,Artamonov2008}:

\begin{eqnarray}
\langle\cos^{2}(\theta,\, t)\rangle & = & \sum_{J,K,M,J'}\langle\psi(t)|\cos^{2}(\theta)|\psi(t)\rangle\label{eq:cos2_wavefunction}\\
 & = & \sum_{J,K,M,J'}c_{J'}^{*}(t)c_{J}(t)\langle J'KM|\cos^{2}(\theta)|JKM\rangle\nonumber 
\end{eqnarray}
where the matrix element can be calculated analytically. The expectation
value can also be written in terms of the axis distribution. For the 1D case this is given by:

\begin{equation}
\langle\cos^{2}(\theta,\, t)\rangle=\intop_{0}^{2\pi}\cos^{2}(\theta)P(\theta,\, t)d\theta\label{eq:cos2_Ptheta}
\end{equation}

Thus $\langle\cos^{2}(\theta,t)\rangle$ can be obtained directly
from the wavefunction, or extracted from the full axis distribution
by projection onto $\cos^{2}(\theta)$. This is just another way of
obtaining a second-order moment of the axis distribution, expressed in a cosine basis.
By use of eqns. \ref{eq:Ptheta_D} - \ref{eq:Ptheta_PL}, selected according to the dimensionality of the problem, the moments obtained in this fashion can also be directly related to the ADMs for any order expectation value $\cos^{n}(\theta)$. Continuing with the 1D case (eqn. \ref{eq:Ptheta_PL}), this yields:

\begin{equation}
\langle\cos^{n}(\theta,\, t)\rangle=\sum_{K}A_{K}(t)\intop_{0}^{2\pi}\cos^{n}(\theta)\mathscr{P}_{K}(\cos(\theta))d\theta\label{eq:cosn_AK}
\end{equation}

The $\langle\cos^{n}(\theta)\rangle$ are therefore a linear combination of the ADMs, weighted by the overlap integral for each $K$. From these considerations, it is clear that the use of $\langle\cos^{2}(\theta,t)\rangle$
as a metric for characterizing an aligned distribution depends on the
exact characteristics of $P(\theta,\, t)$. Distributions with significant
high-order $A_{K,Q}$ may not be well-described by $\langle\cos^{2}(\theta,t)\rangle$
alone. Similarly, the nature of the probe process will dictate whether
higher-order terms are coupled into the observable, providing a second
criterion for the necessity of higher-order terms. Although this
point has been discussed before in the literature (e.g. ref. \cite{Suzuki2005}),
it appears to be the case that most work on aligned distributions
considers only the second-order ($\cos^{2}(\theta,t)$) moment, so demonstration of the effects
of higher-order moments remains a significant motivation in this work.
For non-cylindrically symmetric cases, expectation values of $\cos^{2}$
for other angles are also used to quantify the full 3D alignment \cite{Underwood2005,Artamonov2008},
and recently a single metric for 3D alignment based on this has been
proposed \cite{Makhija2012}; naturally these metrics will similarly
be of best utility for probe process insensitive to higher-order moments
of the distribution.

\begin{figure}
\includegraphics{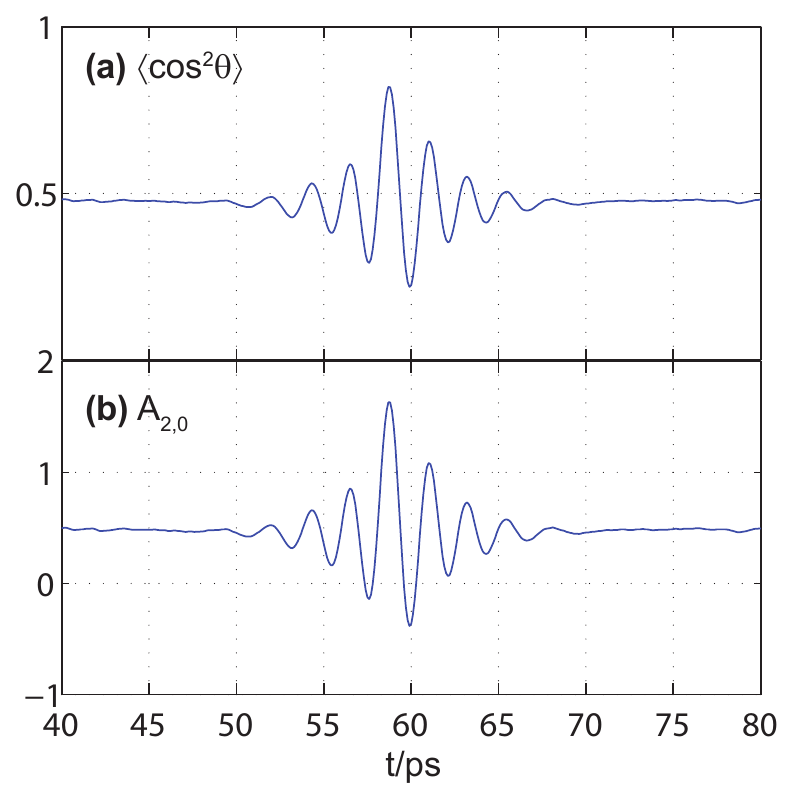}

\caption{Alignment metrics (a) $\langle\cos^{2}(\theta,t)\rangle$ and (b)
$A_{2,0}(t)$ corresponding to the $P(\theta,t)$ distribution shown
in figure \ref{fig:Ptheta_cart_pol}.\label{fig:Alignment-metrics}}

\end{figure}

Continuing the concrete example of a wavepacket calculation for butadiene (treated as a symmetric top),
as sketched above and illustrated in figure \ref{fig:Ptheta_cart_pol},
the corresponding alignment metrics $\langle\cos^{2}(\theta,t)\rangle$
and $A_{2,0}(t)$ are shown in figure \ref{fig:Alignment-metrics}.
Here $\langle\cos^{2}(\theta,t)\rangle$ was calculated directly
from the rotational wavefunction, as defined in eqn. \ref{eq:cos2_wavefunction},
while $A_{2,0}(t)$ was found by fitting the calculated $P(\theta,t)$
to an expansion in spherical harmonics, as defined in eqn. \ref{eq:Ptheta_YKQ}. In this case the temporal response of the two metrics is identical, and they can be regarded as providing effectively equivalent information on the second-order moment of the axis distribution. Full discussion of these results are again deferred to section \ref{sec:Application}.

\subsection{Ionization model}

Here we consider the resulting photoelectron signal from ionization
of an aligned distribution for a variety of cases. We begin with a
full treatment of 1-photon ionization, then generalize to the case
of $N$-photon ionization, and finally discuss ionization via a resonant
intermediate state. All these cases are typical of contemporary experiments
on aligned systems although, to the best of our knowledge, only the
former has been treated in detail in the literature with regard to
ionization of aligned ensembles. The treatment presented here also
includes frame rotations between the aligned frame and the probe frame,
allowing for treatment of polarization-based methodologies such as
angle-resolved ATI \cite{Mikosch2013a}.

In all cases the aligned distribution is assumed to be created via
an intense IR pulse which prepares a broad rotational wavepacket as
described above. The resulting distribution $P(\theta,t)$ is described
in terms of ADMs, i.e. as a geometric parameter, as detailed above.
Ionization is treated within the dipole approximation, hence the ionization
matrix elements have the general form $\langle\Psi^{+};\Psi^{e}|\mu.\mathbf{r}|\Psi^{i}\rangle$,
and describe the dipole coupling of the initial state $\Psi^{i}$
to the final composite state, composed of ion and photoelectron. By
expanding the ionization matrix elements in terms of angular and radial
functions most of the geometric coupling of the ionization matrix
elements can be described analytically, allowing for certain, molecule-independent,
properties \& phenomena to be determined \cite{Dill1976}. 

Treating the distribution geometrically is valid for the ionization
step providing that (a) rotations are separable from the vibronic
wavefunction, hence there is no coupling between rotational dynamics
and other molecular properties; (b) the energy-dependence of the ionization
matrix elements is negligible over the span of rotational levels populated;
(c) that the ionization laser pulse can be treated as constant over
the energy span of the rotational wavepacket, allowing all initially
populated rotational states to be coupled into the ionization. Under
these approximations the angular momentum coupling can be solved analytically
\cite{Stolow2008}. In the case of ionization schemes involving more
than a single photon this stipulation should also hold provided there
are no dynamics in the intermediate state(s) on the time-scale of
rotational motion.%
\footnote{This is essentially the sudden approximation as applied to rotational
motions.%
} For direct $N$-photon ionization via virtual states on the time-scale
of the laser pulse (typically $<$100~fs), this is expected to be
a valid approximation. For the case of ionization via a resonant intermediate
- a typical scheme in pump-probe type experiments investigating excited
state molecular dynamics - this condition may be broken depending
on the time-scales involved. In such cases, an explicit treatment
of the rotational wavepacket on the excited state would be required
to fully account for the evolution of $P(\theta,t)$ on the time-scale
of the experiment, but a sudden-type approximation without inclusion
of these additional dynamics may still provide a reasonable starting
point. Finally, we note that the treatment given here assumes that
the light-matter interaction is perturbative, so does not allow for
field intensity effects. As such, it is not generally valid for strong-field
ionization but, geometrically, should still provide useful insight
to angle-resolved measurements at low computational cost.

\subsubsection{1-photon ionization}

The 1-photon case has been extensively treated in the literature \cite{Underwood2000,Seideman2002,Stolow2008,Ramakrishna2012}.
We recount here the salient details, with a specific focus on the
coupling of the observable to the $A_{K,Q}$, then proceed to determine the properties of specific types of measurement and extend the formalism to the $N$-photon case.

The full photoelectron angular distribution can be most generally
expressed as a multipole expansion (analogous to $P(\theta,\phi,t)$
discussed above, c.f. eqn. \ref{eq:Ptheta_YKQ}):

\begin{equation}
I(\theta,\phi,t)=\sum_{L,M}\beta_{L,M}(t)Y_{L,M}(\theta,\phi)
\end{equation}

Here the polar coordinates reference the laboratory frame (LF), as defined by the probe
pulse (see figure \ref{fig:Example-axis-distributions}(d)), in which
the photoelectron flux as a function of angle and time is measured.%
\footnote{Although omitted here, there is also an energy dependence to the dipole
matrix elements and, hence, to the observable $I(\theta,\phi,t)$.%
} The LF $\beta_{L,M}(t)$ can be written in terms
of the coherent square of the dipole matrix elements: for the ionization
of an aligned ensemble, in the perturbative and dipole approximations,
and assuming that all time-dependence is contained in the axis distribution,
the $\beta_{L,M}(t)$ can be written as \cite{Underwood2000,Stolow2008}:

\begin{widetext}

\begin{eqnarray}
\beta_{L,M}(t) & = & (2L+1)^{1/2}\sum_{P}(-1)^{P}\left(\begin{array}{ccc}
1 & 1 & P\\
p & -p & R
\end{array}\right)e_{-p}e_{-p}^{*}\nonumber \\
 & \mathrm{x} & \sum_{K}\sum_{Q}(2K+1)^{1/2}\left(\begin{array}{ccc}
P & K & L\\
Q-M & -Q & M
\end{array}\right)A_{K,-Q}(t)\nonumber \\
 & \mathrm{x} & \sum_{q,q'}(-1)^{q'}\left(\begin{array}{ccc}
1 & 1 & P\\
q & -q' & q'-q
\end{array}\right)\left(\begin{array}{ccc}
P & K & L\\
q-q' & q'-q & 0
\end{array}\right)\nonumber \\
 & \mathrm{x} & \sum_{l,l'}\sum_{\lambda,\lambda'}(-1)^{\lambda'}(2l+1)^{1/2}(2l'+1)^{1/2}\left(\begin{array}{ccc}
l & l' & L\\
\lambda & -\lambda' & M
\end{array}\right)\left(\begin{array}{ccc}
l & l' & L\\
0 & 0 & 0
\end{array}\right)\nonumber \\
 & \mathrm{x} & (-i)^{l'-l}\sum_{\Gamma,\Gamma'}\sum_{\mu,\mu'}\sum_{h,h'}b_{hl\lambda}^{\Gamma\mu*}b_{h'l'\lambda'}^{\Gamma'\mu'}\boldsymbol{D}_{hl}^{\Gamma\mu*}(q)\boldsymbol{D}_{h'l'}^{\Gamma'\mu'}(q')\label{eq:LF-PAD-t}
\end{eqnarray}

\end{widetext}

The first line of equation \ref{eq:LF-PAD-t} describes the polarization
state of the ionizing radiation; the photon carries 1 unit of angular
momentum with projection $p$ onto the lab frame $z$-axis. For linearly
polarized light aligned with the laboratory frame $z$-axis $p=0$,
hence from the 3-$j$ symbol $P=0,\,2$ and $R=0$. The spherical
tensor components $e_{-p}$ describe the polarization and amplitude
of the ionizing radiation, for the case of linearly polarized light
along the $z$-axis $e_{-p}=e_{0}=e_{z}$ and the term $e_{z}e_{z}^{*}$
can be set to equal unity.

The second \& third lines of equation \ref{eq:LF-PAD-t} describe
the convolution of the molecular frame with the aligned axis distribution,
$P(\theta,t)$, expressed as ADMs. The light field has molecular frame
(MF) projection terms $q$. Terms in $q=0$ thus represent ionizing
light polarized along the MF axis, while $q=\pm1$ terms represent
light polarized perpendicular to the MF axis. If the LF and MF are
coincident then a single value of $q=p$ is selected, while an arbitrary
rotation serves to mix terms in $q$ as the LF polarization axis is
projected onto different MF axes. This mixing (and averaging), due
to the ADMs, is described by the coupling of $P$ and $K$ into the
final multipole moments $L$.

The remaining lines of equation \ref{eq:LF-PAD-t} deal with the photoelectron
and ``molecular'' terms. Here $(l,\lambda)$ represent the photoelectron
partial wave components \cite{Dill1976,messiah}, with (orbital) angular
momentum $l$, and MF projection $\lambda$. The terms $\boldsymbol{D}_{hl}^{\Gamma\mu*}(q)$
represent the symmetrized radial components, with symmetrization coefficients
$b_{hl\lambda}^{\Gamma\mu}$ (see appendix \ref{sub:Symmetry}), of
the (radial) dipole matrix elements for each symmetry-allowed continuum
$\Gamma$ \cite{Stolow2008,reid1991a,Park1996},

\begin{equation}
\boldsymbol{D}_{hl}^{\Gamma\mu}(q)=\langle\Psi^{+};\,\psi_{hl}^{\Gamma\mu,\, e}|\sum_{s}r_{s}Y_{1,q}(\mathbf{\hat{r}}_{s})|\Psi^{i}\rangle\label{eq:r_llam_n}
\end{equation}
where $\psi_{hl}^{\Gamma\mu,\, e}$ are the partial wave components of the photoelectron wavefunction $\Psi^e$ and the summation is over all electrons $s$. These matrix elements
are complex, and may also be written in the form $\boldsymbol{D}_{hl}^{\Gamma\mu}=|\boldsymbol{D}_{hl}^{\Gamma\mu}|e^{-i\eta_{hl}^{\Gamma\mu}}$,
where $\eta$ is the total phase of the matrix element, often called
the scattering phase. The radial matrix elements and phases are
the only part of equation \ref{eq:LF-PAD-t} which are not analytic
functions and, in general, must be determined numerically \cite{Lucchese1982,Arasaki1999}
or from experiment \cite{reid1991,Gessner2002,Hockett2009} for quantitative
understanding of a given system. Symmetry-based arguments can, however,
provide a means of determining which integrals are non-zero, hence
which $(l,\lambda)$ can appear in $\Psi^{e}$. Such considerations
therefore allow for phenomenological, qualitative, or possibly semi-quantitative,
treatments of photoionization for a given molecule, and are discussed
in appendix \ref{sub:Symmetry}.

The effect of the averaging over a distribution of molecular axis
directions is to lose sensitivity in the PADs. In particular, the
observed anisotropy in the LFPAD cannot be more than that arising
from the coupling of the probe photon to the aligned distribution
of molecules, as can be seen from the 3-$j$ term linking terms $P,\, K,\, L$.
This limits $L$ to the range $|P-K|...P+K$ in integer steps. For
instance, if the alignment is prepared by a single pump photon then
a $\cos^{2}\theta$ axis distribution is created, and the only non-zero
alignment parameters are $A_{0,0}$ and $A_{2,0}$. Because $P=0,\,2$
only, the alignment in this case would restrict $\beta_{L,M}(t)$ to
terms with $L=0,\,2,\,4$ (additionally, for cylindrically symmetric
cases, $M=-Q=0$). As the degree of alignment increases higher-order
$K$ terms are required to describe the axis distribution
and the LF ensemble result approaches the true MF \cite{Underwood2000}.
Higher order terms in equation \ref{eq:LF-PAD-t} can be observed,
hence more information is present in the LFPAD and a greater sensitivity
to any property which affects the PADs, e.g. the evolution of the
axis distribution itself, intermediate state dynamics in a pump-probe
experiment, and so on, may be obtained. 

For an angle-integrated measurement (photoelectron yield), integration
over $\{\theta,\,\phi\}$ leaves only the leading term $\beta_{0,0}(t)$,
and angular coherences between partial waves are integrated out of
the measurement. In this case the terms remaining in \ref{eq:LF-PAD-t}
are significantly restricted by the 3-$j$ terms. Allowed terms have
$l=l'$ and $P=K$. For instance, for 1-photon ionization of a distribution
with $P=0,2$, only $A_{K,Q}(t)$ with $K=0,2$ will be coupled to
the ionization yield. Depending on the magnitudes of the parallel
and perpendicular ionization matrix elements, this can result in the
1-photon yield mapping the $A_{2,0}(t)$ 
somewhat directly \cite{Suzuki2005}. 

To make the geometric convolution of the axis distribution and molecular
frame photoionization more explicit, eqn. \ref{eq:LF-PAD-t} can also
be written in the form \cite{Underwood2000}:

\begin{equation}
\beta_{L,M}(t)=\sum_{K,Q}A_{K,-Q}(t)a_{KLM}
\end{equation}
where the $a_{KLM}$ contain all the other terms of eqn. \ref{eq:LF-PAD-t}.
As per eqn. \ref{eq:LF-PAD-t}, the allowed values of $M$ and $Q$
are coupled; for cylindrically symmetric geometries, i.e. probe polarization
parallel to the pump polarization, $Q=M=0$. In this case the photoelectron
yield is given by: 

\begin{eqnarray}
\beta_{0,0}(t) & = & \sum_{K=0,2}A_{K,0}(t)a_{K00}\\
 & = & A_{0,0}(t)a_{000}+A_{2,0}(t)a_{200}\label{eq:B00t-1photon}
\end{eqnarray}
Since the zero-order term $A_{0,0}(t)$ is just the total population
of the rotational states forming the wavepacket (often normalized
to unity), it is a constant in the absence of any population dynamics,
and any time-dependence observed in the ionization yield is due to
the second-order term, as asserted above.

The effect of frame rotations on the $\beta_{L,M}(t)$ can also be
simply expressed using this form, and employing the Wigner rotation
matrix:

\begin{eqnarray}
\beta'_{L,M'}(t;\,\Omega) & = & \sum_{K}\sum_{Q,Q'}D_{Q',Q}^{K}(\Omega)A_{K,-Q}(t)a_{KLM}\label{eq:BOmega_AKQ}\\
 & = & \sum_{K}\sum_{Q'}A'_{K,-Q'}(t)a_{KLM'}
\end{eqnarray}
where the second form follows from the assumption that $Q=0$. Here
$\Omega=\{\Phi,\Theta,\chi\}$ is the set of Euler angles describing
the frame rotation between the alignment field and the probe field,
and the properties in the rotated frame are denoted by primes. The
rotation mixes multipole components $Q$ within a given rank $K$.
Because of the coupling between $Q$ and $M$ (the second 3-$j$ term
in eqn. \ref{eq:LF-PAD-t}) different terms, $M'$, are allowed in
the rotated frame. As discussed above, and illustrated in figures
1 \& 2, such a frame rotation can break the symmetry of an initially
cylindrically symmetric distribution, hence $M'$ may be non-zero
even if $M=0$.

Applying again the stipulations above, the ionization yield for a
cylindrically symmetric distribution ($Q=0$) under a rotation of
$\Theta$ between the alignment and ionization fields, is then given
by:

\begin{eqnarray}
\beta'_{0,0}(t;\,\Theta) & = & \sum_{K=0,2}\sum_{Q'}D_{Q',0}^{K}(0,\Theta,0)A_{K,0}(t)a_{K00}\label{eq:BTheta_AKQ_cyl}\\
 & = & \sum_{K=0,2}\sum_{Q'}d_{Q',0}^{K}(\Theta)A_{K,0}(t)a_{K00}\\
\nonumber & = & A_{0,0}(t)a_{000}+d_{0,0}^{2}(\Theta)A_{2,0}(t)a_{200}\\ 
 &   & +2d_{2,0}^{2}(\Theta)A_{2,0}(t)a_{200}\\
 & = & A_{0,0}(t)a_{000}+(A'_{2,0}(t)+2A'_{2,2}(t))a_{200}
\end{eqnarray}
where $d_{Q',Q}^{K}(\Theta)$ is the reduced Wigner rotation matrix
element, and we have used the identities $d_{0,0}^{0}=1$, $d_{2,0}^{2}=d_{-2,0}^{2}$
\cite{zareAngMom}. This form makes explicit the fact that the frame
rotation mixes additional $Q$ terms into $\beta'_{0,0}(t;\,\Theta)$
as compared to $\beta_{0,0}(t;\,\Theta=0)$ (eqn. \ref{eq:B00t-1photon},
see also figure \ref{fig:ADMs-under-rotation}). Therefore, in general,
a measurement of $\beta'_{0,0}(t;\,\Theta\neq0)$ will not map $A_{2,0}(t)$
as directly as $\beta'_{0,0}(t;\,\Theta=0)$.

A measurement of the photoelectron yield as a function of $\Theta$
will therefore have the general form (using eqn. 3.93 from Zare \cite{zareAngMom}):

\begin{widetext}
\begin{equation}
I(\Theta;\, t)=\beta'_{0,0}(\Theta;\, t)=A_{0,0}(t)a_{000}+\left(\frac{4\pi}{5}\right)^{\frac{1}{2}}A_{2,0}(t)a_{200}Y_{2,0}^{*}(\Theta,0)+2\left(\frac{4\pi}{5}\right)^{\frac{1}{2}}A_{2,0}(t)a_{200}Y_{2,2}^{*}(\Theta,0)\label{eq:BTheta_AKQ_cyl_YLM}
\end{equation}
\end{widetext}

Experimentally, this corresponds to a measurement of the photoionization
yield as a function of polarization geometry (defined by $\Theta$,
as illustrated in fig. \ref{fig:Example-axis-distributions}(d)).
Measurements of this form may be made for all $\Theta$, yielding
the polarization-angle-resolved ionization yield as a quasi-continuous
function, or compared at selected $\Theta$ to provide ``transient
anisotropy'' measurements, for example the standard formulation (which
compares yields at $\Theta=0$ and $\Theta=\pi/2$) has been explored
with respect to ionization \cite{Schalk2011}. Clearly, a polarization-resolved
measurement of this form may be expected to display relatively complex
angular structure, with up to 4 lobes on the interval $0\leq\Theta\leq2\pi$
due to the summation of $Y_{2,0}$ and $Y_{2,2}$ terms, despite the
fact that only the second-order moment of the axis distribution is
invoked. This serves to illustrate how even an apparently simple experimental
measurement may respond in a more complex fashion than anticipated
to an aligned ensemble.

\subsubsection{Multi-photon ionization\label{sub:Multi-photon}}

Formally, a direct multi-photon ionization process constitutes a ladder
of transitions through virtual states. In terms of the decomposition
of this ladder of transitions into dipole matrix elements for each
successive photon absorption - a direct extension of the 1-photon
case discussed above - the complexity rapidly grows, and may be further
complicated by near-resonances with real bound states. A full treatment
of such cases for atomic ionization has been given by Bebb \& Gold
\cite{Bebb1966}, including the derivation of an effective $N$-photon
matrix element to reduce the complexity of the problem.%
\footnote{For more details see also Lambropoulos, Maragakis and Zhang \cite{Lambropoulos1998},
who discussed many of the general issues in multi-photon ionization,
including details of a range of formal treatments. Examples of the
treatment of $N$-photon absorption for small $N$, in which all matrix
elements are treated explicitly using a similar formalism to that
employed in this work, can be found in, for example, refs. \cite{Bain1984,Dubs1988,Docker1988}.
A more recent treatment, including discussion of perturbative vs.
non-perturbative regimes, can be found in ref. \cite{Toffoli2012}.%
} Here we take a similar approach and consider treating the ionization
as an effective 1-photon transition in which the photon angular momentum
is large. The ionization would then have a form essentially as eqn.
\ref{eq:LF-PAD-t}, but with the photon carrying $N$ units of angular
momentum. Such a treatment should allow for at least a qualitative
picture of the directionality of the ionization, and in particular
the polarization-angle-resolved ionization yields $I(\Theta,t)$,
with a clear link to the multipoles involved in the molecular axis
alignment \& ionization process. 

The immediate result is that $P$ can take many more values than in
the 1-photon case. For a cylindrically symmetric distribution eqn.
\ref{eq:BTheta_AKQ_cyl} becomes:

\begin{equation}
\beta'_{0,0}(t;\,\Theta)=\sum_{K=0...2N}\sum_{Q'}D_{Q',0}^{K}(0,\Theta,0)A_{K,0}(t)a_{K00}\label{eq:B00t-theta_N-photon}
\end{equation}
Hence higher-order ADMs may be coupled to the observed signal, according
to the value of $N$. In practice one would expect, therefore, to
observe more structure in the alignment trace (temporal signal) $\beta'_{0,0}(t;\,\Theta)$
for a given $\Theta$, due to the coupling of larger $K$ into the
signal. Similarly, more angular structure may be observed in the angle-resolved
yield $I(\Theta,t)$ and the PADs, $I(\theta,t;\,\Theta)$, recorded
for a given polarization geometry. Because $l_{max}$, the maximum
photoelectron angular momentum, will also grow with $N$, it is likely
that high-order ADMs are always coupled to such high-order processes,
regardless of the exact details of the photoionization matrix elements.

Physically, one can understand the higher-order angular momenta as
signifying a more directional ionization event. For example, in the
limit of tunnel ionization, the outgoing electron is confined to a
narrow angular spread by the shape of the tunnel, and one can envisage
a jet of electron flux centred on the laser polarization axis.%
\footnote{Although, to reiterate a point already noted above, since a perturbative
geometric treatment of the type shown here will not take intensity
effects into account, for processes involving laser fields beyond
the perturbative limit ($\gtrsim10^{11}$~Wcm$^{-2}$) it is best
considered as a phenomenological comparator for the angular dependence
of different polarization geometries and/or photon orders at a fixed
intensity. Furthermore, this picture is somewhat of a simplification
for complex systems, see for instance refs. \cite{Murray2011,Spanner2013}.%
} In the framework of angular momentum theory, this is exactly equivalent
to a process with high angular momentum. This is the same concept
as discussed in section \ref{sub:Axis-distribution-moments}, where
it was seen that contributions from larger angular momenta $K$ give
rise to a sharper, or more directionally localised, axis distribution.
Ultimately, the angular-dependence of any observable, expressed in
terms of an expansion in angular functions, can contain very high-order
terms for processes which are highly directional.

\subsubsection{Ionization via resonant intermediate states\label{sub:2-photon-excitation}}

The $N$-photon case discussed above assumed that all intermediate
states were virtual and, consequently, did not provide any restrictions
on the ionization. A distinctly different case arises when there is
a single, or multiple, resonant intermediate state(s), because bound-bound
transitions carry strict selection rules. In the context of this discussion
we are concerned with the directionality, or polarization, of the
bound-bound transition, as determined by the electronic dipole selection
rules. Here we discuss the simplest case of a single bound-bound transition
at the 1-photon level prior to ionization - a 1+$N$ resonantly enhanced
multiphoton ionization (REMPI) process. The geometric formalism presented
could readily be extended to more complex processes.

In the case of a 1-photon bound-bound transition the initially prepared
$P(\theta,t)$ will be raised by an additional power of $\cos^{2}(\theta)$
for a parallel transition, or $\sin^{2}(\theta)$ for a perpendicular
transition, where the transition direction is defined by the symmetry
of the bound states involved and the dipole operator. This can be
considered purely geometrically, and the transition amplitudes are
not required to model this process assuming only a single transition
is allowed (or multiple transitions are resolved and can be considered
independently). Strictly, such a transition will change the composition
of the rotational wavepacket; however, under the assumption that there
are no dynamics on the excited state (ionization is effectively instantaneous/rapid
following excitation as compared to rotations as discussed above)
- hence no rotational wavepacket propagation need be taken into account
- then a geometric treatment is valid. Furthermore, the 1-photon absorption
contains coupling terms which change $J$ by $0,\pm1$, so will only
minimally affect the envelope of a broad wavepacket. 

In a purely geometric treatment, the axis distribution following a
parallel transition is given by:

\begin{eqnarray}
P'_{\parallel}(\theta,\, t) & = & \cos^{2}(\theta)P(\theta,\, t)\label{eq:Ptheta_resonant_para}
\end{eqnarray}
and for the perpendicular case by:

\begin{equation}
P'_{\perp}(\theta,\, t)=\sin^{2}(\theta)P(\theta,\, t)\label{eq:Ptheta_resonant_perp}
\end{equation}

The eqns. above can be written more explicitly in terms of the initial
ADMs; for the parallel case we have: 

\begin{eqnarray}
P'_{\parallel}(\theta,\, t) & = & \cos^{2}(\theta)\sum_{K,Q}A_{K,Q}(t)Y_{K,Q}(\theta,\phi)\\
 & \propto & \frac{1}{3}(Y_{2,0}(\theta,0)+1)\sum_{K,Q}A_{K,Q}(t)Y_{K,Q}(\theta,\phi)
\end{eqnarray}
where the $\propto$ arises because some normalization factors have
been neglected. Similarly, for the perpendicular case we have: 

\begin{eqnarray}
P'_{\perp}(\theta,\, t) & = & \sin^{2}(\theta)\sum_{K,Q}A_{K,Q}(t)Y_{K,Q}(\theta,\phi)\\
 & \propto & Y_{2,2}(\theta,0)\sum_{K,Q}A_{K,Q}(t)Y_{K,Q}(\theta,\phi)
\end{eqnarray}
For cylindrically symmetric cases ($Q=0$) and single photon ionization
($K=0,\,2$) these eqns. simplify further to:

\begin{widetext}
\begin{eqnarray}
P'_{\parallel}(\theta,\, t) & \propto & A_{0,0}(t)(1+Y_{2,0}(\theta,0))+A_{2,0}(t)(Y_{2,0}(\theta,0)+Y_{2,0}(\theta,0){}^{2})
\end{eqnarray}
and

\begin{eqnarray}
P'_{\perp}(\theta,\, t) & \propto & A_{0,0}(t)Y_{2,2}(\theta,0)+A_{2,0}(t)Y_{2,0}(\theta,0)Y_{2,2}(\theta,0)
\end{eqnarray}
where $Y_{0,0}(\theta,0)$ has been assumed to be normalized to unity.

Hence, eqn. \ref{eq:B00t-1photon} becomes:

\begin{equation}
\beta_{0,0}^{\parallel}(t)=A_{0,0}(t)a_{000}(1+Y_{2,0}(\theta,0))+A_{2,0}(t)a_{200}(Y_{2,0}(\theta,0)+Y_{2,0}(\theta,0)^{2})\label{eq:B00t-2photon-REMPI}
\end{equation}
for the parallel case and, for the perpendicular case:

\begin{equation}
\beta_{0,0}^{\perp}(t)=A_{0,0}(t)a_{000}Y_{2,2}(\theta,0)+A_{2,0}(t)a_{200}Y_{2,0}(\theta,0)Y_{2,2}(\theta,0)\label{eq:B00t-2photon-REMPI-perp}
\end{equation}
\end{widetext}

As expected, in both cases higher-order angular moments appear in
the observable, although only the second-order moment of the originally
prepared $P(\theta,\, t)$ appears in the final equations. These equations
show explicitly how resonant multi-photon ionization processes may
respond to aligned ensembles in relatively complex ways, due both
to the additional photon angular momentum coupled into the ionization
and the angle-dependence of the resonant step. As was the case for
frame rotations, such measurements respond to the $P(\theta,t)$ but
in a less-than-direct manner, and this response is highly dependent
on the nature of the probe process.

\section{Application\label{sec:Application}}

Following from the theory outlined above, we explore the response
of single and multi-photon ionization to molecular alignment for angle-integrated
and angle-resolved measurements. We first consider the ADMs in detail,
and discuss general features which might be expected for any molecular
ensemble. Time-resolved ionization for a typical aligned ensemble,
for various angle-integrated and angle-resolved measurements, is then
discussed. Specifically, we consider the limiting cases of a 1-photon
probe and 1+1' REMPI scheme, both of which are common experimental
techniques, with respect to both the axis distribution and the ionization
matrix elements. For the multi-photon case we investigate the effect
of $N$ on the observables, as a function of the axis distribution,
for a specific ionization process. These results should provide a
general guide to experimentalists working with aligned distributions,
since the order of the terms coupled into a given observable can significantly
affect the experimentally measured quantities.

\subsection{Axis distribution moments}

\subsubsection{General features\label{sub:General-features-ADMs}}

\begin{figure}
\includegraphics{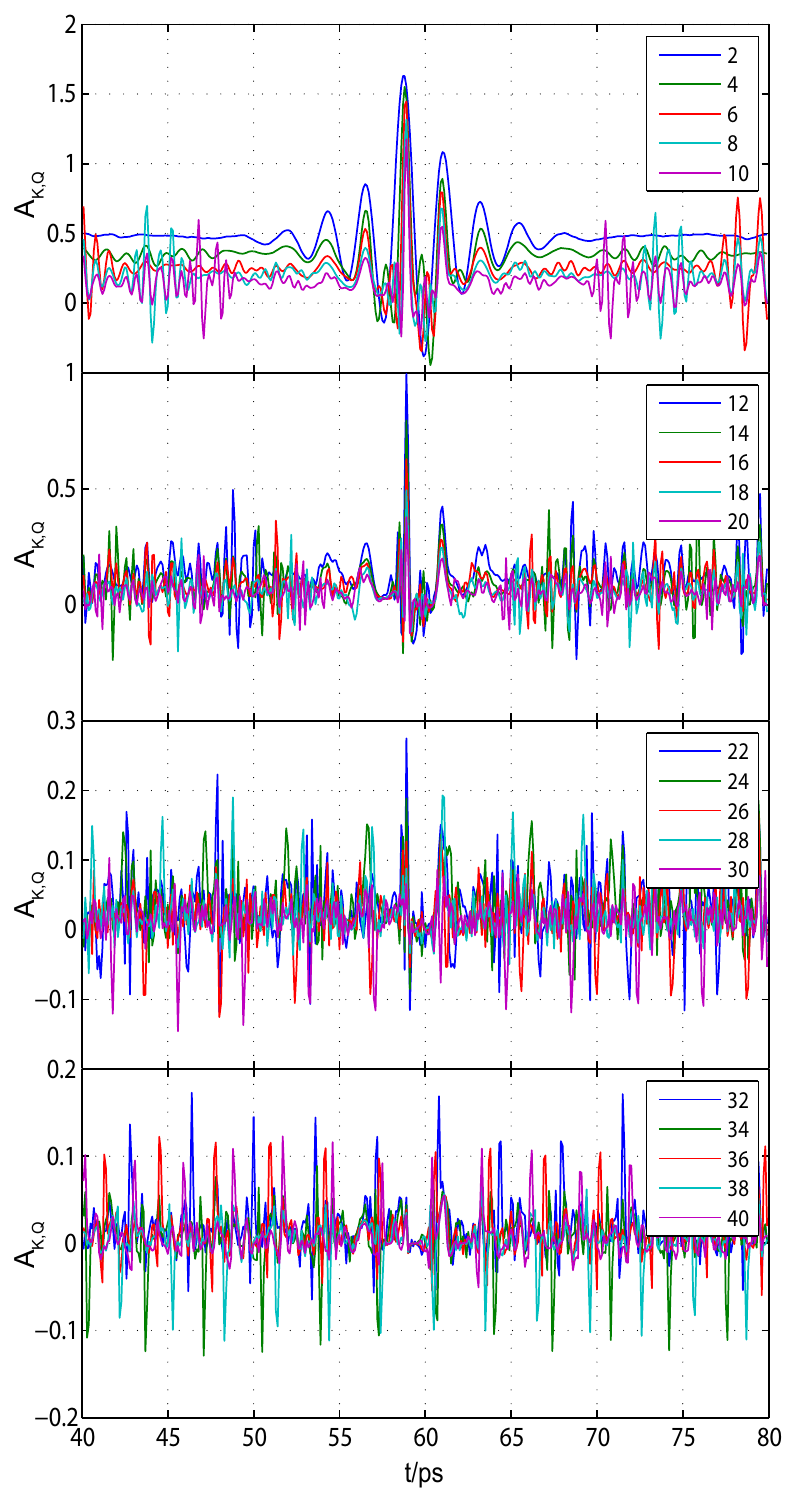}

\caption{ADM expansion for calculated $P(\theta,t)$ as shown in figure \ref{fig:Ptheta_cart_pol}.
For clarity each plot shows only five $A_{K,0}(t)$ terms, and the
expansion in $K$ is only shown up to $K=40$.\label{fig:full-ADM-expansion}}

\end{figure}

Calculation results for $P(\theta,t)$, in the region of the half-revival
of a rotational wavepacket calculated for butadiene under typical experimental conditions, have already been illustrated in figure
\ref{fig:Ptheta_cart_pol}. The axis distributions were mapped in
both Cartesian and polar space, and also expressed in terms of the
associated second-order alignment metrics, $\langle\cos^{2}(\theta,t)\rangle$
and $A_{2,0}(t)$, in figure \ref{fig:Alignment-metrics}. As discussed
above, and clear from the complex angular structure visible in figure
\ref{fig:Ptheta_cart_pol}, a full description of $P(\theta,t)$
requires higher-order moments of the axis distribution to be taken
into account. Figure \ref{fig:full-ADM-expansion} illustrates this
point with the temporal evolution of all ADMs up to $K=40$ for this
axis distribution. As already noted with respect to the $P(\theta,t)$
distributions, in this case it is clear that the degree of alignment
is high, and many terms in $K$ are significant in the $A_{K,Q}(t)$
expansion around the half-revival. Away from the peaks of the revival
the degree of alignment is still high, due to an effective DC contribution
to $P(\theta,t)$, although there are also high-frequency temporal
modulations, particularly visible at the poles of the $P(\theta,t)$
distribution. 

The $A_{K,Q}(t)$ which parametrise the full $P(\theta,t)$ distribution
also show complex behaviour, as expected from the form of $P(\theta,t)$.
Since we are concerned with the coupling of the aligned distribution
- parametrized as a set of $A_{K,Q}(t)$ - into an observable, we
discuss here the evolution of the axis distribution in terms of the
$A_{K,Q}(t)$ rather than the underlying rotational wavepacket. We
also focus on the comparison of the higher-order terms with $K=2$,
since this corresponds to the most often used metric of alignment.

At the main peak of the alignment, $t\sim59$~ps, all of the terms
up to $K\simeq24$ peak, reflecting the maximal axis alignment obtained
at the half-revival. The widths of the features reveal a more complex
temporal dependence of different order terms, with a narrowing of
the revival peak at higher $K$. Immediately before and after this
feature, at around $t\sim58$ and $t\sim60$~ps respectively, the
traces show additional satellite features appearing at higher $K$.
The satellite features around the half-revival have the effect of
blurring out the anti-alignment features, relative to $K=2$, hence
would potentially reduce the observable contrast of the revival for
an observable sensitive to higher-order terms as compared to a low-order
observable. In general, it is clear that one would expect observables
which couple to different order terms to exhibit different revival
contrast and markedly different temporal evolution. Finally, it is
worth noting that the frequency content of the distribution also scales
with $K$ (this is a direct consequence of terms with high $\Delta J$
preferentially coupling into higher-order moments of the distribution).
A high-order measurement away from the revival peak may, therefore,
be difficult to interpret in terms of the expected (typically low-order)
response and the complex temporal response may even be dismissed as
experimental noise.%
\footnote{There may also be a slight increase in numerical noise in the calculations
for higher-order terms, although the smooth and explicable behaviour
of the $A_{K,Q}(t)$ as a function of $K$ implies this does not have
a significant effect.%
}

As compared with $\langle\cos^{2}(\theta,t)\rangle$, there is significantly more temporal structure in the higher-order $K$ terms which is not coupled into $\langle\cos^{2}(\theta,t)\rangle$. This is clear from comparison of figure \ref{fig:Alignment-metrics}, which shows the direct correspondence between $\langle\cos^{2}(\theta,t)\rangle$ and $A_{2,0}(t)$. This follows from the overlap integral of eqn. \ref{eq:cosn_AK} which, for $n=2$, will be most significant for $K=2$. Furthermore, the first few terms in $K$ have similar temporal response to $K=2$, and smaller magnitudes. The net result is that any contributions to the $\langle\cos^{2}(\theta,t)\rangle$ line-shape from these ADMs will have only a minimal effect on the overall temporal profile, and the $\langle\cos^{2}(\theta,t)\rangle$ metric can be considered as essentially equivalent to $A_{2,0}(t)$ in this case, as stated earlier (sect. \ref{sec:metrics}) without explanation.

It may be concluded that, in general, a highly-aligned ensemble will
contain high-order $A_{K,Q}(t)$ (effectively by definition), regardless
of the precise details of the molecule under study. An appreciation
for the coupling of this ensemble into the measurement is, therefore,
essential for an understanding of experimental results. In particular,
analysis of results based solely on $\langle\cos^{2}(\theta,t)\rangle$
will generally not be sufficient for interpretation of experimental
data, even at a phenomenological level, beyond the simplest single-photon
ionization yield probe scheme.

\subsubsection{Resonant transitions\label{sub:Resonant-transitions}}

\begin{figure}
\includegraphics{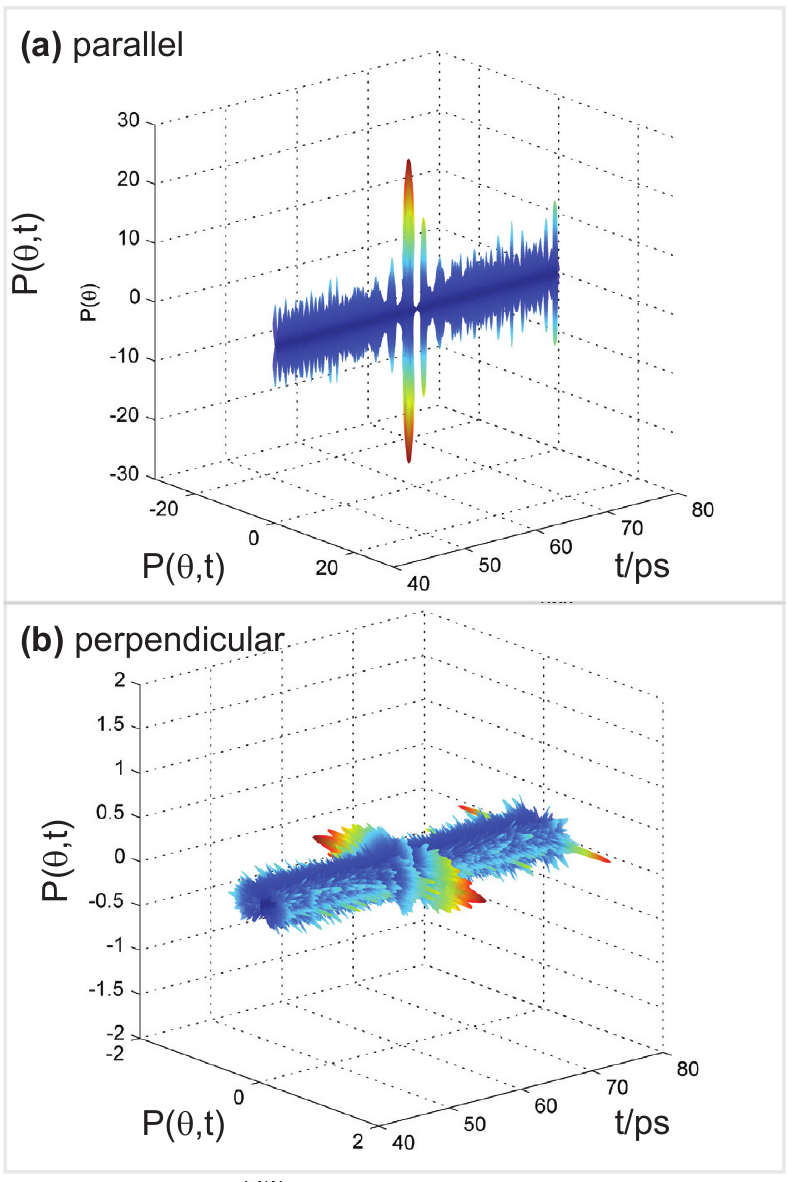}

\caption{$P'(\theta,\, t)$ following (a) parallel and (b) perpendicular bound-bound
transitions. These distributions are based on the initial distribution,
$P(\theta,t)$, as shown in figure \ref{fig:Ptheta_cart_pol}.\label{fig:Pt-states}}
\end{figure}

\begin{figure}
\includegraphics{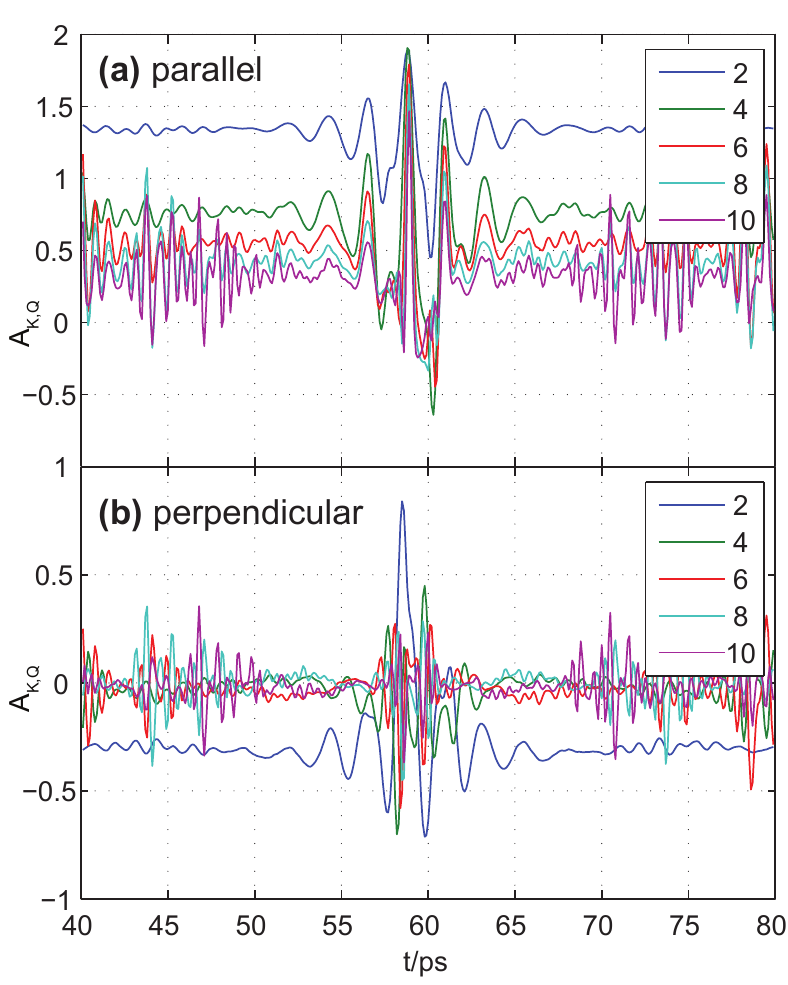}

\caption{$A_{K,Q}(t)$ following 1-photon excitation (shown for $K\leq10$).\label{fig:AKQ_gs_ex}}
\end{figure}

In order to treat the case of 1+1' REMPI, we consider distributions
$P_{\parallel}'(\theta,\, t)$ and $P_{\perp}'(\theta,\, t)$ following
parallel and perpendicular bound-bound transitions respectively, obtained
via eqns. \ref{eq:Ptheta_resonant_para} \& \ref{eq:Ptheta_resonant_perp}.
These distributions are shown in figure \ref{fig:Pt-states}. The
parallel transition enhances the alignment, with a slight sharpening
of the distribution at the poles (c.f. fig. \ref{fig:Ptheta_cart_pol}),
as might be intuitively expected from the form of eqn. \ref{eq:Ptheta_resonant_para}.
However, the perpendicular transition significantly changes the axis
alignment. This is also an intuitive result: essentially the perpendicular
transition couples preferentially to axes aligned in the $(x,\, y)$
plane, thus addressing only population away from the poles ($z$-axis)
of the initial distribution. Consequently the peaks in the distribution
are shifted both spatially (into the ($x,\, y$) plane) and temporally,
with the peak in the $(x,y)$ plane of $P_{\perp}'(\theta,\, t)$
corresponding with the anti-alignment feature in the initial $P(\theta,\, t)$.
As a consequence of the coupling, the high-order oscillations in the
$(x,\, y)$ plane are, in effect, amplified.

Figure \ref{fig:AKQ_gs_ex} illustrates these effects in terms of
the ADMs and shows the $A_{K,Q}(t)$, up to $K=10$, for both cases.
In particular the effect of a parallel transition is to raise the
second-order moment ($K=2$) of the distribution, while a perpendicular
transition will decrease 
the second-order moment. The
temporal profiles are also affected, with additional modulations appearing
leading to significant temporal asymmetry in the $K=2$ 
traces, particularly in the case of a perpendicular transition. Essentially
the effect of the transition is to mix higher-order terms into a given
$K$; for example the $A_{2,0}(t)$ trace for $P_{\parallel}'(\theta,\, t)$
has characteristics of both $A_{2,0}(t)$ and $A_{4,0}(t)$ of the
initial $P(\theta,t)$ - this can be seen by comparison of fig. \ref{fig:AKQ_gs_ex}
and fig. \ref{fig:full-ADM-expansion} (top panel). Because single
photon ionization is only sensitive to $K=0,2$ terms of the aligned
ensemble, as discussed above, the resonant excitation step has the
effect of allowing $K=4$ 
terms from the initial $P(\theta,t)$ to contribute to this observable since
these are now mixed into the $A_{2,0}(t)$ ADMs for the excited state
distribution $P'(\theta,\, t)$.

Again, in general one can conclude from this discussion that for more
complex ionization processes higher-order terms become necessary to
understand and interpret experimental results. In particular the perpendicular
case illustrates how a process with strong selection rules (i.e. a
highly directional response to the molecular axis alignment) acts
as a strong filter on the initially prepared axis distribution. In
the most general case of multiphoton ionization, where multiple resonances
may be accessed sequentially or via different competing ionization
pathways, it is clear that a very complex $P'(\theta,\, t)$ may be
created which has little obvious correspondence to the initially prepared
$P(\theta,\, t)$ and, without some appreciation of the underlying
probe process, the temporal response of the observable may be inexplicable.

\subsection{Angle-integrated observables}

\subsubsection{1-photon ionization yields\label{sub:Limiting-cases}}

We first consider some limiting cases in order to investigate the
response of the time-resolved photoelectron yield, $\beta_{0,0}(t)$,
to $P(\theta,t)$. We consider the case of (a) 1-photon ionization
from the rotationally-excited ground state, (b) 1-photon ionization
following a parallel ($\cos^{2}(\theta)$) resonant excitation and
(c) 1-photon ionization following a perpendicular ($\sin^{2}(\theta)$)
resonant excitation. The calculated $P(\theta,t)$ and the extracted
$A_{K,Q}(t)$ for these cases have already been shown in figures \ref{fig:Ptheta_cart_pol},
\ref{fig:full-ADM-expansion}, \ref{fig:Pt-states} and \ref{fig:AKQ_gs_ex},
and discussed above. In all cases we assume that the probe (ionization)
pulse polarization is parallel to the pump (alignment) pulse polarization,
i.e. $\Theta=0$. For the excitation step the laser pulse polarization
is assumed to be either parallel or perpendicular to the alignment pulse polarization,
as appropriate for the excitation, and its effect is treated geometrically
(no temporal evolution of the wavepacket in the excited state) as
discussed in sect. \ref{sub:2-photon-excitation}.

\begin{figure}
\includegraphics{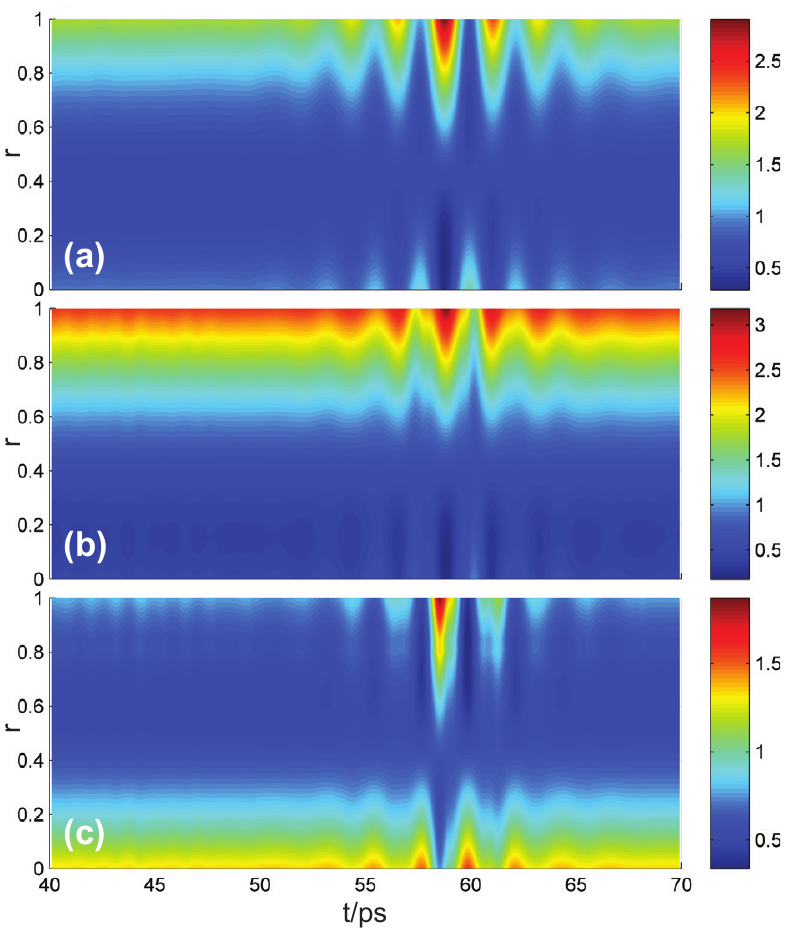}

\caption{Limiting cases for 1-photon ionization of an aligned distribution.
Calculations are based on (a) calculated $P(\theta,t)$, (b) $P'_{\parallel}(\theta,t)$
for an excited state populated via a parallel 1-photon transition
and (c) $P'_{\perp}(\theta,t)$ for an excited state populated via
a perpendicular 1-photon transition. Ionization matrix elements are
set such that $r$=0 corresponds to a purely perpendicular ionization
event, and $r$=1 a purely parallel ionization.\label{fig:Limiting-cases-1-photon}}
\end{figure}

The dipole matrix elements in eqn. \ref{eq:LF-PAD-t} typically represent
unknown quantities. In order to explore limitations on the ionization
yield we consider the yield as a function of the parallel vs. perpendicular
ionization cross-sections ($\sigma_{\parallel}$ and $\sigma_{\perp}$).
Specifically, in the case of butadiene - which is used here as an
exemplar system - ionization of the $S_{2}$ excited state in $C_{2h}$
symmetry leads to allowed continuum symmetries of $A_{g}$ and $B_{g}$
character for parallel and perpendicular ionization events respectively (see appendix \ref{sub:Symmetry} for details). The effect of varying
the ratio of the ionization matrix elements is shown in figure \ref{fig:Limiting-cases-1-photon}.
Here the matrix elements were set as 
\[
\boldsymbol{D}_{hl}^{\Gamma\mu}\equiv\sum_{q}\boldsymbol{D}_{|m|l}^{\Gamma}(q)=r\boldsymbol{D}_{|m|l}^{A_{g}}(0)+(1-r)\boldsymbol{D}_{|m|l}^{B_{g}}(\pm1)
\]
and all symmetry-allowed terms, up to $l_{max}=4$, were included
and set to unity. The matrix elements were further re-normalised such
that $|\sum_{l,|m|}b_{|m|l|m|}^{A_{g}}\boldsymbol{D}_{|m|l}^{A_{g}}(0)|^{2}=1$
and $|\sum_{l,|m|}b_{|m|l|m|}^{B_{g}}\boldsymbol{D}_{|m|l}^{B_{g}}(\pm1)|^{2}=1$.
Hence $r=1$ corresponds to the $A_{g}$ continuum only, and a purely
parallel ionization event in the molecular frame, while $r=0$ corresponds
to the $B_{g}$ continuum and a purely perpendicular ionization.
Interestingly, the same type of calculation can also be used to extract
the ratio of the matrix elements from experimental data by fitting
of the matrix elements \cite{Ramakrishna2012}, although, due to the
structure of eqn. \ref{eq:LF-PAD-t}, and as emphasised in eqn. \ref{eq:B00t-1photon},
the $\beta_{0,0}(t)$ are sensitive only to the amplitudes of each
$l$ (although can contain cross-terms in $\lambda$), so phase information
between different $l$ is not defined, and cannot be obtained, from
this observable alone. 

These results - figure \ref{fig:Limiting-cases-1-photon}(a) - show quantitatively
a number of features which might intuitively be expected. The ionization
yield is sensitive only to $A_{2,0}(t)$, as shown in eqn. \ref{eq:B00t-1photon}.
For the purely parallel case ($r=1$), the $A_{2,0}(t)$ is mapped
faithfully (this is shown in detail in figure \ref{fig:B00_N-order}(a)),
while the purely perpendicular case ($r=0$) shows an inverted trace,
essentially mapping $\langle\sin^{2}(\theta,t)\rangle$. Additionally,
the total yield is decreased in the perpendicular case due to the
geometry - $P(\theta,t)$ is heavily peaked along the $z$-axis as
shown in fig. \ref{fig:Ptheta_cart_pol}, so the axis distribution
geometrically favours $q=0$ transitions which are parallel to the
molecular axis, hence aligned near to the LF $z$-axis. Between these
limits, $0<r<1$, the ionization yield is less sensitive to the axis
distribution, and even shows regions of no sensitivity where the balance
of parallel and perpendicular components is such that there is no
observable modulation in the yield. In other words, these are regions
where there is no angular dependence for a second-order observable
due to the equal magnitudes of parallel and perpendicular ionization
cross-sections. The import of this is that, for a given molecule,
it is possible that an observable sensitive to only the lowest order
ADM shows no alignment dependence, despite the presence of an aligned
distribution (c.f. measuring at magic angle in order to cancel out
$\cos^{2}(\theta)$ terms). Since this depends on a molecular property
- the details of the ionization matrix elements - it is generally
out of the control of the experimentalist, and may not be anticipated
\emph{a priori} unless the molecule under study is already well-characterized.
Such a situation becomes less likely as higher-order moments are coupled
to the observable.

Ionization yields following 1-photon excitation, figure \ref{fig:Limiting-cases-1-photon}(b) \& (c), show similar behaviour,
with the main difference appearing due to changes in the temporal
line-shapes of $P'(\theta,t)$ as discussed in section \ref{sub:Resonant-transitions}.
The parallel yield following a parallel excitation, i.e. ionization
of the distribution $P'_{\parallel}(\theta,t)$ with $r=1$, is larger for
all $t$ (relative to case (a)), and is not modulated as strongly
around the revival feature; conversely the perpendicular yield is
reduced relative to case (a). The opposite is seen following a perpendicular
excitation, i.e. ionization of the distribution $P'_{\perp}(\theta,t)$
(case (c)), which enhances the yield in the case of perpendicular
ionization, and reduces it for parallel ionization. In both cases
(b) \& (c) this effect is due to the coupling of $K=4$ terms from
the initial axis distribution into the observable, as discussed above
(sect. \ref{sub:2-photon-excitation}), and consequently also results
in more complex temporal evolution of the signal, as would be expected
from the form of the $A_{K,Q}(t)$ (figure \ref{fig:full-ADM-expansion}).
Despite this additional term, there are still values of $r$ where
the yield shows very little sensitivity to the ADMs, although the
width of this region of $r$-space appears significantly reduced in
the perpendicular case.

\subsubsection{Multi-photon ionization yields\label{sub:Multi-photon-ionization-yields}}

\begin{figure}
\includegraphics{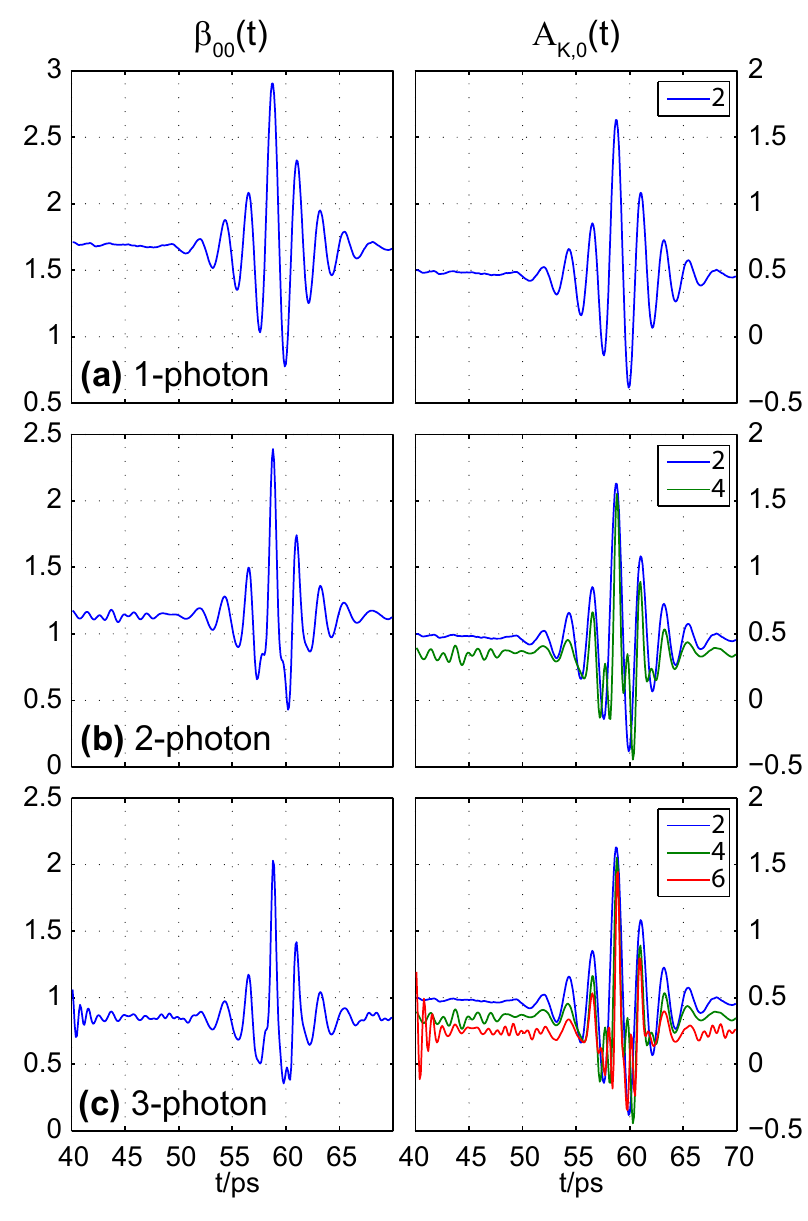}

\caption{$\beta_{0,0}(t;\Theta)$ for $N$-order ionization processes. (Left
column) $\beta_{0,0}(t;\,\Theta=0)$ for $N=1-3$ and (right column)
$A_{K,0}(t)$ for $K_{max}=2N$. \label{fig:B00_N-order}}
\end{figure}

In the multiphoton case, as detailed in sect. \ref{sub:Multi-photon},
the expectation is for higher-order $A_{K,Q}(t)$ terms to become
significant as the photon order of the process increases. This behaviour
is illustrated in figure \ref{fig:B00_N-order} for $N=1-3$, and
compared directly with the contributing $A_{K,Q}(t)$. In these calculations
$r=1$, hence the ionization is purely parallel and, as for the 1-photon
case above, the probe and alignment pulse polarization are parallel
($\Theta=0$). The line-outs show how the total ionization yield drops
as a function of $N$, and also how the line-shapes change as couplings
with larger $K$ become allowed. Comparison of $\beta_{0,0}(t)$ with
$A_{K,Q}(t)$ clearly shows that the line-shapes contain contributions
from $A_{K,Q}(t)$ up to $K_{max}=2N$ (see also eqn. \ref{eq:B00t-theta_N-photon}):
there is a narrowing of the features observed over the revival, and
an increased complexity and temporal asymmetry to the line-shape as
a function of $N$ (equivalently as a function of $K_{max}$), as
already observed in the ADMs (sect. \ref{sub:General-features-ADMs}).
The oscillations away from the revival also increase, again as expected
from the $A_{K,Q}(t)$ traces. Naturally, the exact coupling will
depend on both $r$ (as illustrated for the 1-photon case in the preceding
section) and $\Theta$ (see below), but these general features will
likely be apparent to some degree in an $N$-photon observable.

\subsection{Angle-resolved observables}

\subsubsection{Angle-resolved ionization yields}

Angle-resolved measurements, based on the response of the ionization
yield to the probe polarization, are given by $\beta_{0,0}(\Theta,t)$
as defined in equation \ref{eq:B00t-theta_N-photon}. This is essentially
the same observable as discussed above - the total ionization yield,
angle-integrated with respect to the photoelectron - except that measurements
are made as a function of pump-probe polarization geometry, i.e. the
angle $\Theta$, as illustrated in figure \ref{fig:Example-axis-distributions}(d).

\begin{figure}
\includegraphics{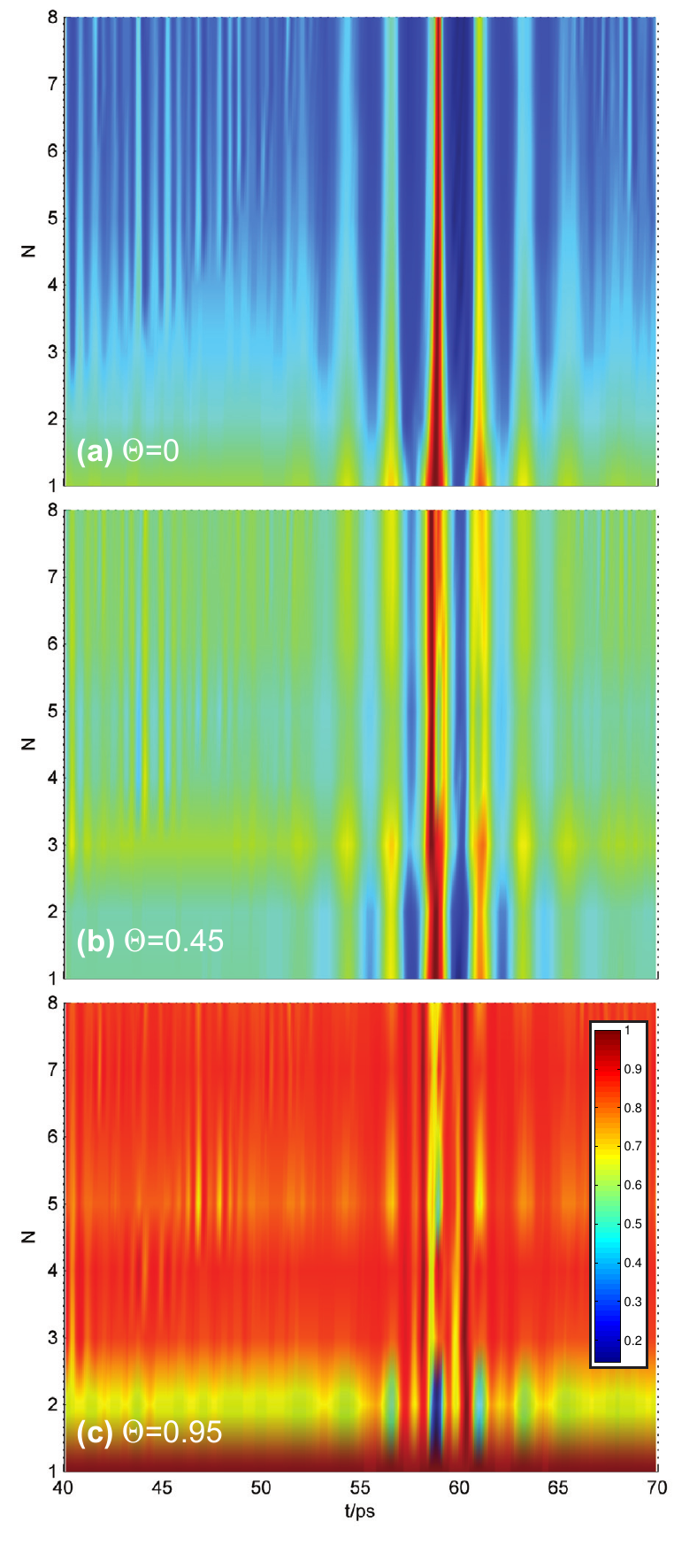}

\caption{$\beta_{0,0}(t;\Theta)$ for $N$-order ionization processes. (a)$\beta_{0,0}(t;\,\Theta=0)$
for $N=1-8$, (b) as (a) but $\Theta=0.45$~rad, (c) as (a) but $\Theta=0.95$~rad.
The surface plots are re-normalised to the peak of the signal for
each $N$ to emphasize the temporal behaviour.\label{fig:B00_N-order_surfs}}
\end{figure}

Figure \ref{fig:B00_N-order_surfs} shows surface plots for of $\beta_{0,0}(t;\,\Theta)$
for $N=1-8$ and $\Theta=0,\,0.45,\,0.95$~rad (0$^{\circ}$,~26$^{\circ}$,~54$^{\circ}$)
respectively (all other parameters were the same as those used in
sect. \ref{sub:Multi-photon-ionization-yields}); the data is re-normalised
for each $N$ to emphasize the changes in the line-shapes independently
of the total yield. For ease of comparison, panel (a) shows the same
results as figure \ref{fig:B00_N-order} for $N=1-3$, and the narrowing
of the features with $N$, as discussed above, is again very clear.
In cases (b) and (c) very different surfaces are observed, as expected
from eqn. \ref{eq:B00t-theta_N-photon}. The line-shapes are more
complex, and this is particularly apparent in the splitting of the
main feature observed for $N>3$ in fig. \ref{fig:B00_N-order_surfs}(b).
As would also be expected, the 54$^{\circ}$ case shows reduced contrast 
as the aligned distribution is now rotated by close to $\pi/4$ from
the probe pulse, hence the probe now maps a combination of $\langle\cos^{K}(\theta)\rangle$
and $\langle\sin^{K}(\theta)\rangle$. For $N=3$ in fig. \ref{fig:B00_N-order}(b)
and $N=4$ in fig. \ref{fig:B00_N-order}(c) a decreased sensitivity
to alignment is observed. This is, presumably, due to the most geometrically
significant $A_{K,Q}(t)$ term (most strongly coupled according to
eqn. \ref{eq:B00t-theta_N-photon}) coming close to its magic angle,
but may also be due to the distinct $\Theta$-dependence of different
order terms leading to regions where contrast is washed out. This
is essentially the same effect discussed above, and shown in figure
\ref{fig:B00_N-order}(c) for $\Theta=0$ and $N=3$, in which the
$t$-dependence, rather than $\Theta$-dependence, of different terms
led to a decrease in revival contrast. 

\begin{figure}
\includegraphics{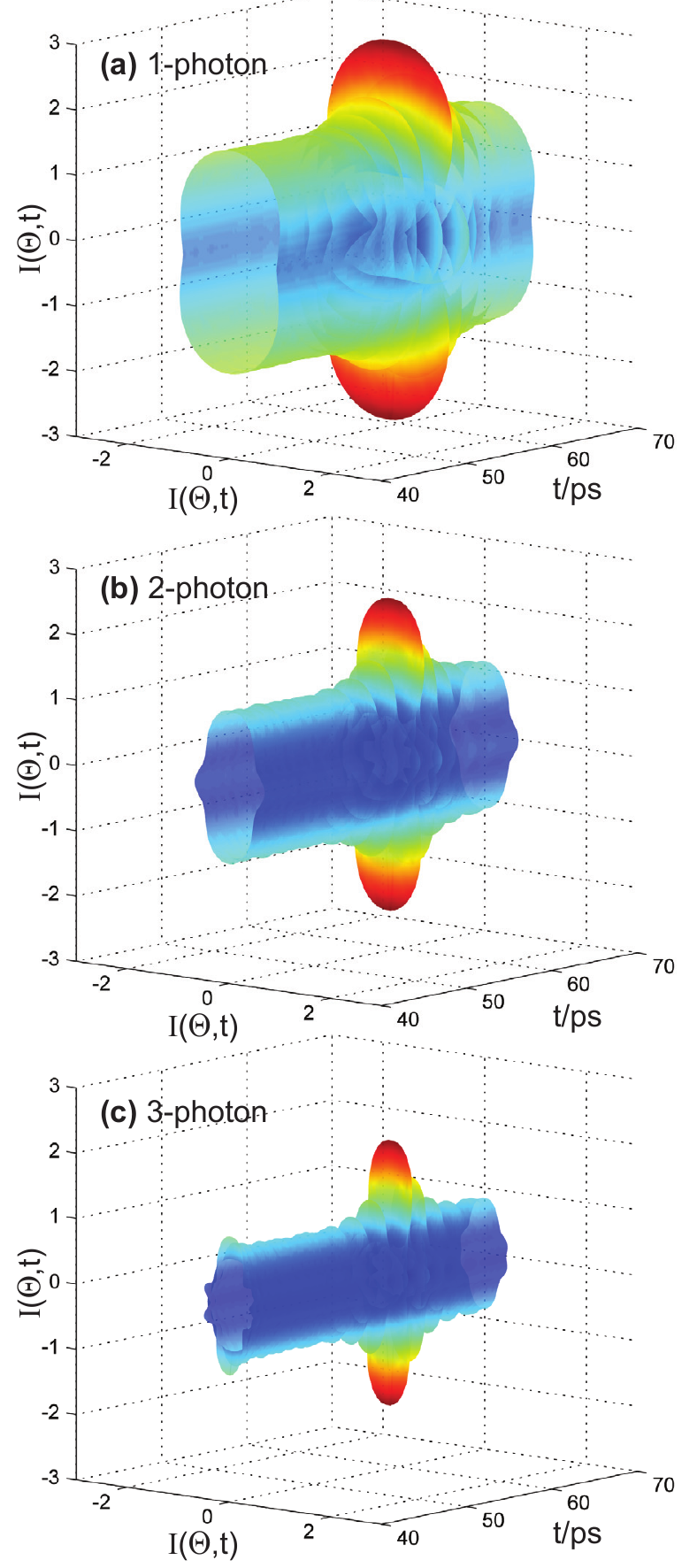}

\caption{Polar plots of $I(\Theta,\, t)$ for $N=1-3$.\label{fig:It-theta_N-photon}}

\end{figure}

To further visualize this behaviour, figure \ref{fig:It-theta_N-photon}
shows full surfaces of $I(\Theta,\, t)$ for $N=1-3$ in polar form.
These plots show more clearly the narrowing of the observed distributions
at the poles as $N$ increases. Comparison with fig. \ref{fig:Ptheta_cart_pol}(b)
shows that the narrowing of the equator, as well as the increased
spatial and temporal complexity, of higher $N$ cases approach the
full $P(\theta,\, t)$ surface - a direct result of higher $A_{K,Q}(t)$
terms becoming coupled into the observable, hence an increase in information
content or fidelity with respect to the axis distribution. In this
context, the most direct and detailed experimental measure for mapping
an aligned distribution should be a multi-photon probe of high order,
with terms up to $K_{max}=2N$ present in the observable in a - relatively
- transparent manner, provided that the $\Theta$-dependence of the
probe process (i.e. the ionization matrix elements) is well-defined
as in this illustration.

\subsubsection{PADs}

As was the case for the $N$-photon ionization yields discussed above,
PADs can be considered as high-order observables with many $\beta_{L,M}(t)$
contributing to the observable. Hence, as indicated in eqn. \ref{eq:LF-PAD-t},
many $A_{K,Q}(t)$ may be mapped by the PAD. In terms of phenomenology,
the extra sensitivity of the observed PAD to both the amplitudes and
phases of the ionization matrix elements makes it difficult to choose
realistic representative values for use in limiting case calculations.
In this case we simply use the same limiting cases as above, i.e.
$A_{g}+B_{g}$ continuum functions with amplitudes set to unity and
phases set to zero, and explore the observables as a function of $r$. Generally, one could hope to obtain the matrix elements from experimental measurements of PADs if the alignment is
known \cite{Suzuki2005,Suzuki2006,Ramakrishna2012}; conversely, one
could also use experimentally-measured PADs to map alignment in the
case where the ionization matrix elements are known \cite{Reid2000,Ramakrishna2013}.

\begin{figure}
\includegraphics{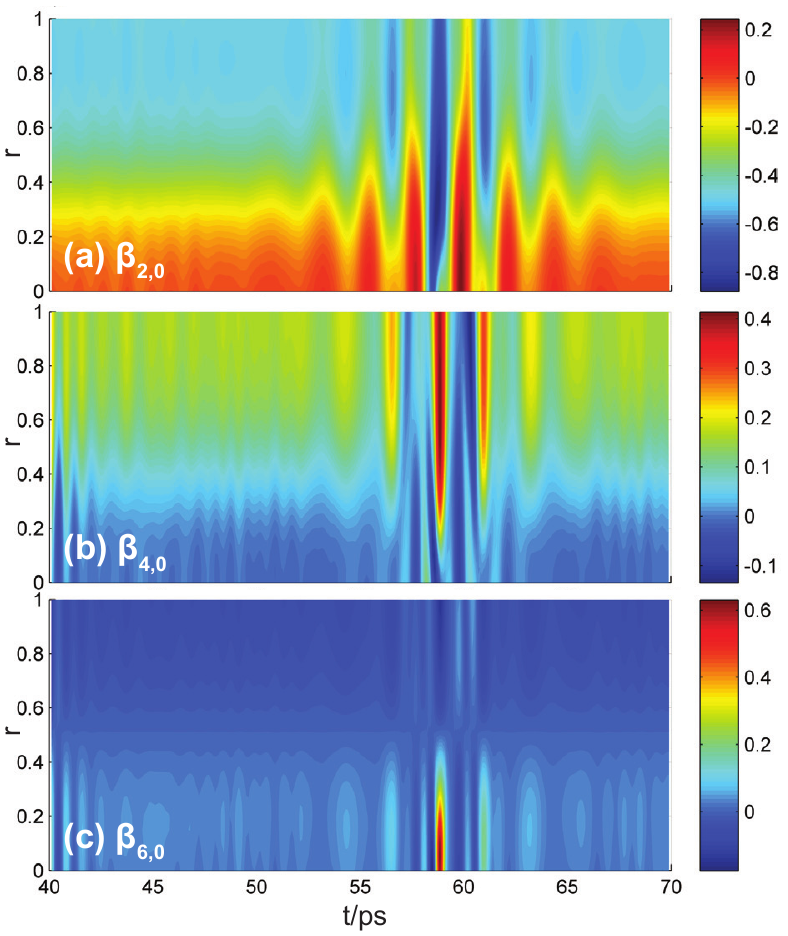}

\caption{$\beta_{L,M}(t;\,\Theta=0)$ for 1-photon ionization, with (a) $L=2$,
(b) $L=4$ and (c) $L=6$; $M=0$ in all cases. Ionization matrix
elements are set such that $r$=0 corresponds to a purely perpendicular
ionization event, and $r$=1 a purely parallel ionization, as per
results already presented for $L=0$ (ionization yield) in figure
\ref{fig:Limiting-cases-1-photon}.\label{fig:BLM_1-photon} \textcolor{red}{}}

\end{figure}

Figure \ref{fig:BLM_1-photon} shows examples of $\beta_{L,M}(t;\,\Theta=0)$
calculated for these example cases, as a function of $r$ as per the
results discussed in section \ref{sub:Limiting-cases} and shown in
fig. \ref{fig:Limiting-cases-1-photon}. Similarly to the $I(\Theta,\, t)$
for $N$-photon ionization discussed above, higher-order $L$ terms
couple to higher-order $K$, resulting in more complex line-shapes
which map primarily different $A_{K,Q}(t)$ as a function of $L$.
One immediate result is that, with the exception of $r=0.5$ and $L=6$,
there are no regions totally insensitive to the axis alignment in
this case, in contrast to the $L=0$ results shown in figure \ref{fig:Limiting-cases-1-photon}(a).
Additionally, the higher-order terms are sensitive to the phases of
the ionization matrix elements, so will display a much enhanced sensitivity
to molecular properties (e.g. vibronic dynamics) as compared to the yields ($L=0$) alone.
Another interesting observation in this particular case is that the
temporal peak in the $\beta_{2,0}(t)$ (at the half-revival) does not
move significantly with $r$, although it does narrow and shift slightly.
This is very different to the ionization yield, where the observable
essentially mapped $A_{2,0}(t)$ for $r=1$ (parallel ionization) but
was inverted for $r=0$ (perpendicular ionization), hence the half-revival
feature appeared out of phase for $r=0$ as compared to $r=1$ (see
fig. \ref{fig:Limiting-cases-1-photon}(a)). However, for the $\beta_{4,0}(t)$
very different behaviour is observed, and there is a significant temporal
shift in the main feature as a function of $r$, although the switch
is not abrupt and does not occur at $r=0.5$ as is the case for the
yield (and as one might intuitively expect). Naturally, the specific
details of these types of behaviour are highly dependent on the ionization
matrix elements, but in general it is clear that a more complex temporal
response is expected regardless of the exact details of the ionization
matrix elements.

\begin{figure}
\includegraphics{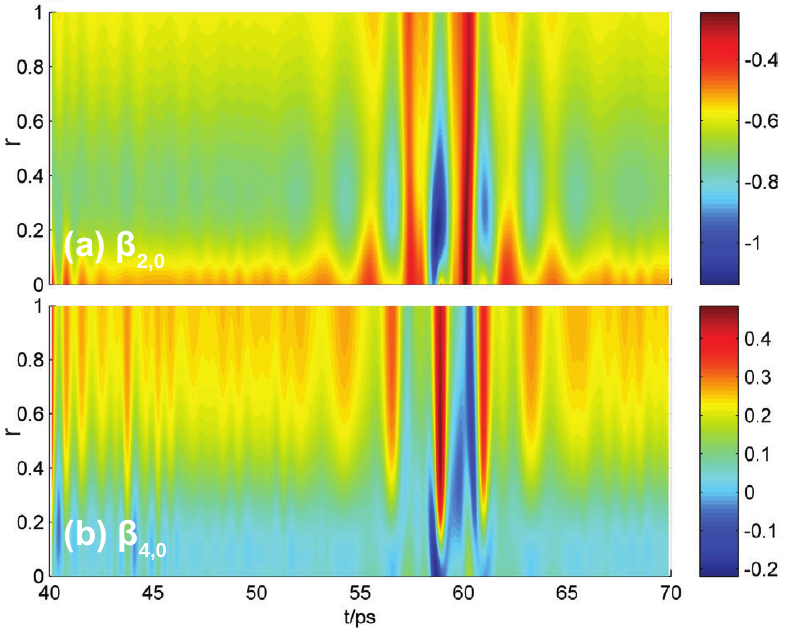}

\caption{$\beta_{L,M}^{\parallel}(t;\,\Theta=0)$ for 1-photon ionization, following parallel
excitation, with (a) $L=2$, (b) $L=4$; $M=0$ in all cases. Ionization
matrix elements are set such that $r$=0 corresponds to a purely perpendicular
ionization event, and $r$=1 a purely parallel ionization, as per
results already presented in figures \ref{fig:Limiting-cases-1-photon}
and \ref{fig:BLM_1-photon}.\label{fig:BLM_1plus1_para}}

\end{figure}

\begin{figure}
\includegraphics{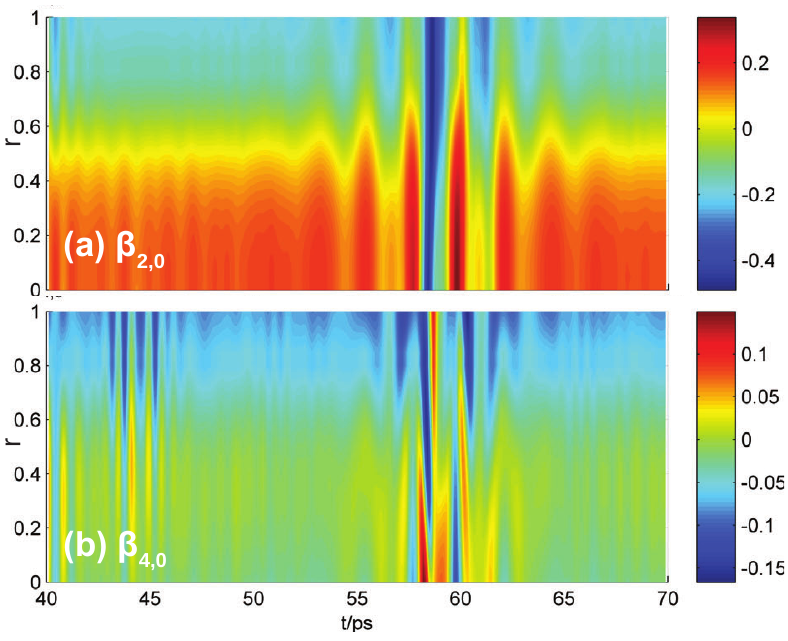}

\caption{$\beta_{L,M}^{\perp}(t;\,\Theta=0)$ for 1-photon ionization, following perpendicular
excitation, with (a) $L=2$, (b) $L=4$; $M=0$ in all cases. Ionization
matrix elements are set such that $r$=0 corresponds to a purely perpendicular
ionization event, and $r$=1 a purely parallel ionization, as per
results already presented in figures \ref{fig:Limiting-cases-1-photon}, \ref{fig:BLM_1-photon} and \ref{fig:BLM_1plus1_para}.\label{fig:BLM_1plus1-1_perp}}
\end{figure}

To further emphasize the sensitivity of the PADs to the ADMs, figures
\ref{fig:BLM_1plus1_para} and \ref{fig:BLM_1plus1-1_perp} show selected
$\beta_{L,M}$ ($L=2$ and $L=4$) for ionization following 1-photon
absorption, for a parallel (i.e. axis distribution given by $P'_{\parallel}(\theta,t)$)
and perpendicular (i.e. axis distribution given by $P'_{\perp}(\theta,t)$)
absorption; in the following discussion we denote the $\beta_{L,M}(t)$
correlated to these cases as $\beta_{L,M}^{\parallel}(t)$ and $\beta_{L,M}^{\perp}(t)$
respectively. As expected, all of the $\beta_{L,M}$ respond to the
change in the ADMs following absorption, and are much more sensitive
to the details of the axis distribution than the yields alone (sect.
\ref{sub:Limiting-cases}). In particular:
\begin{itemize}
\item Higher-order terms become more strongly coupled. This is apparent
from, for instance, the appearance of strong modulations in the $\beta_{2,0}(t)$
around 40~-~45~ps, far from the main half-revival feature, and
the increase in contrast of such features in the $\beta_{4,0}$.
\item The range \& magnitudes of the $\beta_{L,M}$ change. Interestingly,
for the set of matrix elements used here, $\beta_{2,0}^{\parallel}(t)$ has
only negative values for all $(t,r)$, while the sign of $\beta_{2,0}^{\perp}(t)$
changes at $r\sim0.5$ (except at the half-revival feature). In both
cases the average value of $\beta_{2,0}(t)$ for a given $r$ is significantly
different from ionization of the initial distribution $P(\theta,t)$;
the range of $\beta_{L,M}$ values is similar for ionization of $P(\theta,t)$
and $P'_{\parallel}(\theta,t)$, although offset, but much reduced for $P'_{\perp}(\theta,t)$
(this is particularly evident for $\beta_{4,0}^{\perp}(t)$). Experimentally
this would mean less significant changes in the time-resolved PADs
would be observed in the latter case.
\item The position of the temporal maxima, as noted above, move only slightly
as a function of $r$ for $\beta_{2,0}(t)$, but do move significantly
for $\beta_{4,0}(t)$. Additionally, for $\beta_{2,0}^{\parallel}(t)$, the
width and magnitude of the temporal maxima at the half-revival is
almost constant for all $r$; this is quite distinct from the other
cases.
\item For $\beta_{4,0}^{\perp}(t)$ the line-shapes over the half-revival
are more complex for all $r$, this is distinct from the behaviour
observed for $\beta_{4,0}^{\parallel}(t)$ and all $L=2$ cases, for which
the half-revival features remain qualitatively similar in temporal
complexity to the unexcited case. This change therefore reflects both
the significant change in the ADMs for $P'_{\perp}(\theta,t)$ (see
fig. \ref{fig:AKQ_gs_ex}) and the enhanced coupling to higher-order
ADMs for $L=4$ as compared to $L=2$.
\end{itemize}
These specific features indicate how complex the response of the $\beta_{L,M}(t)$
may be in any given case, with the temporal response reflecting both
the ionization matrix elements and the ADMs. However, we again emphasize
that one can begin to build some intuition on how these observables
may respond phenomenologically to the ADMs and the experimental configuration
in general terms, even if the precise details are molecule dependent.
In this regard it is clear that one might expect a strong response
to molecular alignment in all $\beta_{L,M}(t)$ as compared to the
ionization yield (even in cases where the yield is insensitive to
alignment), and that higher-order terms will contain higher-frequency
components due to the coupling to higher-order ADMs.

This mapping has been discussed extensively by Seideman and co-workers
(e.g. refs. \cite{Suzuki2005,Ramakrishna2012}), including the possibility
of extracting ionization matrix elements via fitting experimental
data in this case. From an experimental perspective, measuring PADs
may therefore be useful for characterizing alignment and ionization
dynamics, but requires a rather involved analysis due to the high
information content and complexity of the coupling \cite{Suzuki2006}.
However, the benefit of such measurements is precisely this high information
content, so in some cases this effort is worthwhile. Similarly, in
experiments where PADs are used as probes for other molecular properties
(for a recent review of applications, see ref. \cite{Reid2012}),
it is clear that the effect of rotational dynamics must be carefully
considered precisely due to this complexity.

\section{Discussion \& Conclusions}

In general, from an experimental perspective, one can draw several
qualitative conclusions from the results presented here, some obvious
and some not so. Firstly, regardless of the degree of alignment achieved
in a specific case, there is always the possibility that a given observable
is insensitive to the alignment. In such cases one would often conclude
that the experimental set-up is flawed in some way; however, as demonstrated
in figure \ref{fig:Limiting-cases-1-photon}, at least in the 1-photon
case, there is a reasonable chance that the sensitivity of the photoelectron
yield to alignment is small, or non-existent; for higher-order observables
this is less probable, but contrast over the features of a revival
may still be poor. Secondly, an observable sensitive to high-order
terms may appear highly temporally structured, which could be inexplicable
or even be attributed to experimental noise when considered in terms
of expectations based on typical $\langle\cos^{2}(\theta,t)\rangle$
line-shapes but, if reproducible, is likely a valid result due to
coupling of higher-order ADMs as illustrated in figure \ref{fig:full-ADM-expansion}.
Thirdly, high-order observables are required in order to map aligned
distributions in detail, and multi-photon ionization is probably the
cleanest probe to use to achieve the goal of mapping such distributions
via ionization measurements. 

Finally, following directly from these points, it is worth highlighting
again that the low-order metrics in wide-spread use to describe aligned
distributions (i.e. $\langle\cos^{2}(\theta,t)\rangle$) are of limited
utility for a detailed description of the aligned distribution - and,
consequently, the observable - in any case where higher-order terms
are coupled to the observable under study. Experimentally, although
one may not be interested in the details of the rotational wavepacket
per-say, optimisation of molecular axis alignment and appreciation
of the effects it may have on an observable are certainly prerequisites
to an optimal measurement, and for obtaining results which can be
qualitatively, or even quantitatively, interpreted in terms of the
molecular behaviour under study; clearly, even broad expectations
about how a given signal may look are useful in this regard. These
conclusions are also especially relevant to experiments aimed at measuring
properties approaching the molecular frame, since the coupling of
the alignment moments of the distribution may be different from the
true molecular frame result even in the case of a high degree of alignment,
and for time-resolved experiments where the time-scale of the evolution
of the rotational distribution relevant to the experiment will depend
on which alignment moments couple to the observable. 

As a concrete example of this latter point, consider PADs obtained
via a time-resolved pump-probe methodology, measured with the aim
of studying the excited state molecular dynamics from a ``fixed-in-space''
molecule \cite{Bisgaard2009,Hockett2011,Wang2014}. In this case one must be careful to make pump-probe measurements
on time-scales over which the alignment can be considered static,
hence the temporal evolution of the rotational wavepacket will not
play a role in the observable. However, what this time-scale is will
depend on both the observable (i.e. whether higher-order terms play
a role, see fig. \ref{fig:full-ADM-expansion}) and the details of
the rotational wavepacket prepared, since a narrower rotational wavepacket
will have broader temporal features (combined with a lower degree
of alignment). This time-scale, in some cases, may be $\ll\,1$~ps,
which is typical of the time-scales of (vibronic) molecular dynamics
investigated in such pump-probe measurements, but is usually assumed
to be ``safe'' with respect to the time-scale of rotational wavepacket
evolution. Without such considerations misleading conclusions are
likely with, for example, alignment-mediated signal decays interpreted
as state lifetimes, or changes in angle-resolved observables interpreted
purely in terms of vibrational or electronic wavepacket evolution.
Unfortunately such considerations do nothing to make alignment experiments
easier, but do promise that more detailed and precise measurements
can be made.

For completeness we reiterate that the treatment presented herein assumes a perturbative ionization regime (laser intensity $\lesssim10^{11}$~Wcm$^{-2}$), which in practice may be broken by the laser fields required to drive the high-order multi-photon processes of the kind suggested above for mapping aligned distributions. In such cases there may be modification of the rotational wavepacket during the ionization process, as well as the possibility of other strong-field effects; such considerations will naturally be very much molecule and laser wavelength dependent \cite{Lezius2001}. In general it should however be possible to drive few-photon processes, particularly in the UV, with perturbative fields, and some recent examples illustrating typical experimental conditions which drove various multi-photon processes in this regime can be found in refs. \cite{Vredenborg2008, Hockett2013, Wilkinson2014}. Other recent work, ref. \cite{Hockett2014}, has demonstrated the possibility of combining perturbative and non-perturbative treatments at different photon-orders as one method of efficiently incorporating intra-pulse dynamics driven by an IR field in the moderate intensity regime ($10^{12}-10^{13}$~Wcm$^{-2}$) into photoionization calculations, hence presents a possible means to extend a geometric multi-photon treatment to the non-perturbative regime.

In this work various observables pertaining to single or few-photon
ionization have been considered but, more generally, the same conclusions
apply to other types of measurement, such as Coulomb explosion, high-harmonic
generation \cite{Ramakrishna2013} and X-ray diffraction \cite{Pabst2013},
as well as more traditional measurements such as fluorescence. Of
particular relevance in this regard is the recent article from Ramakrishna
\& Seideman \cite{Ramakrishna2013}, which explicitly considers rotational
wavepacket imaging via different observables (Raman-induced polarization
spectroscopy, 1-photon ionization and high-harmonic generation) as
a function of pump-probe geometry, so is highly complementary to the
observables considered in this work. Simply put, without a detailed
understanding of the couplings involved in a measurement one cannot
hope to understand the details of either the prepared rotational wavepacket
or the relation of the observable to the aligned distribution. This
is an obvious conclusion, but is often ignored in experimental analysis
- even at the phenomenological level - for reasons of simplicity.
A recent illustration of the utility of a more complete experimental
analysis is given in ref. \cite{Lock2012}, where the observation
of high-order rotational revivals in the high-harmonic signal from
an aligned ensemble provided a way to determine the maximum continuum
electron angular momentum, and further analysis also allowed the determination
of the relevant matrix elements.

To summarize, in this work we have considered the coupling of highly-structured
molecular axis distributions, typical of contemporary experiments
utilizing strong IR pulses to prepare broad rotational wavepackets,
into various types of photoionization measurement. The treatment highlighted
the geometric complexity of the axis distributions created, and the
role of the probe interaction in terms of the geometric coupling of
the observable to the axis distribution moments. Insight into the
response of the observables was discussed in general terms, providing
a phenomenology for a range of photoionization-based measurement schemes.
Most generally, at a phenomenological level, this treatment indicates
the types of complex behaviours which might be expected from \emph{any}
measurement technique which couples to high-order axis distribution moments of an aligned molecular ensemble.

\begin{acknowledgments}

We are extremely grateful to Christer Z. Bisgaard for providing mature
code for the rotational wavepacket calculations used in this work.
We thank Rune Lausten, Jochen Mikosch, Michael Spanner and Albert
Stolow for helpful discussions on this topic and manuscript.

\end{acknowledgments}

\appendix		

\section{Numerics\label{sub:Numerics}}

In the results detailed above, calculation of the axis alignment was
made using code developed by C.Z. Bisgaard \cite{Bisgaard2006}, as
part of the work of H. Staplefeldt's group. In these calculations
the light-matter interaction is treated via a time-dependent Schrödinger
equation (TDSE) formalism, in which the set of coupled differential
equations are solved numerically by an adaptive-step Crank-Nicholson
algorithm. This treatment requires knowledge of the (static) polarizabilities
and the rotational constants of the molecule of interest, and the
rotational temperature to determine the initial Boltzmann population
of $|JKM\rangle$ states.

The Hamiltonian for the interaction with the laser pulse is given
by \cite{Bisgaard2006}:

\begin{equation}
\hat{H}(t)=\hat{H}_{rot}+\hat{V}(t)=B\hat{J}^{2}+(A-B)\hat{J}_{z}^{2}-\frac{E(t)^{2}}{4}(\Delta\alpha\cos^{2}\theta+\alpha_{\perp})
\end{equation}
where $A$ and $B$ are rotational constants, $E(t)$ is the (time-dependent)
electric field, $\Delta\alpha$ is the difference between the parallel
($\alpha_{\parallel}$) and perpendicular ($\alpha_{\perp}$) polarizabilities,
and $\hat{J}$ and $\hat{J}_{z}$ are the usual rotational operators
for total rotational angular momentum and its projection onto the
$z$-axis respectively.

After the laser pulse the $c_{J}$ coefficients, that is the populations
of the rotational states comprising the full rotational wavepacket,
are fixed and field-free evolution of the wavepacket is determined
analytically by eqn. \ref{eq:cjt}. The axis distribution at time
$t$ is then calculated as per eqn. \ref{eq:ptheta} which, making
use of eqn. \ref{eq:psiJKM}, can be written as \cite{Bisgaard2006}:

\begin{widetext}
\begin{equation}
P(\Theta,t)=\sum_{J}|c_{J}(t)|^{2}(J+\frac{1}{2})(d_{-M,-K}^{J}(\Theta))^{2}+\sum_{J<J'}2\mathrm{Re}(c_{J}(t)c_{J'}^{*}(t))(J+\frac{1}{2})^{\frac{1}{2}}(J'+\frac{1}{2})^{\frac{1}{2}}d_{-M,-K}^{J}(\Theta)d_{-M,-K}^{J'}(\Theta)
\end{equation}
\end{widetext}
where $d_{-M,-K}^{J'}(\Theta)$ are the (reduced) Wigner rotation
matrix elements.

In the calculations the upper limit for $J$ was set by the approximate
scaling law determined empirically \cite{Bisgaard2006}:

\begin{equation}
J_{max}=-0.35I^{2}+9.2I+30
\end{equation}
where $I$ is the pulse (peak) intensity (in units of TWcm$^{-2}$).

Numerical results from a simplified version of this code, calculating
only $\langle\cos^{2}(\theta,t)\rangle$ have previously been tested
for a variety of cases \cite{Bisgaard2006,Mikosch2013}. In this work
the numerics of the calculation of the full $P(\theta,t)$ calculation
were tested against the existing code by comparison of the $\langle\cos^{2}(\theta,t)\rangle$ parameters, i.e. by making use of equations \ref{eq:cos2_wavefunction}
for ``direct'' and \ref{eq:cos2_Ptheta} for ``indirect'' or geometric
calculations. This procedure was also tested for direct
vs. indirect calculation of $\langle\cos^{4}(\theta,t)\rangle$, providing confidence in the reliability
of all other higher-order terms which were only extracted geometrically.

As noted above, in this work we considered butadiene as our
exemplar system. The relevant molecular properties are given in
table \ref{tab:Molecular-parameters}, which provides the literature values and the symmetrized values used in our calculations, which treat the molecule as a symmetric top. In this case, with the $B$ and $C$ rotational constants within 10, this should be a reasonable approximation but, naturally, the numerical results illustrated here will not show any effects associated with asymmetric top rotational wavepacket dynamics \cite{Holmegaard2007}.
Typical experimental conditions were used in our calculations, with I = 5 TWcm$^{-2}$, $\tau$ = 400~fs, $T_{r}$ = 2~K (rotational temperature). These conditions were chosen to correspond
to recent experimental work on butadiene (which will be discussed in a later publication \cite{hockett2014b}). Extensive studies of rotational temperature effects and intensity averaging were not carried out in this case but, in general, it is expected that both effects will lead to a reduction in the maximum alignment, and a smoothing out of the temporal profile.

\begin{center}
\begin{table}
\begin{centering}
\begin{tabular}{|c|c|c|}
\hline 
Property & Literature & Calculation (symmetrized)\tabularnewline
\hline 
\hline 
A & 1.3903772(6)~cm$^{-1}$ & 1.3903772~cm$^{-1}$\tabularnewline
\hline 
 & 41.6831~GHz & 41.6831~GHz\tabularnewline
\hline 
B & 0.1478868(2)~cm$^{-1}$ & 0.1408~cm$^{-1}$\tabularnewline
\hline 
 & 4.4335~GHz & 4.2211~GHz\tabularnewline
\hline 
C & 0.1336949(3)~cm$^{-1}$ & -\tabularnewline
\hline 
 & 4.0081~GHz & -\tabularnewline
\hline 
$\alpha_{zz}$ & 12.82 & 12.82\tabularnewline
\hline 
$\alpha_{xx}$ & 6.34 & 5.73\tabularnewline
\hline 
$\alpha_{yy}$ & 5.12 & -\tabularnewline
\hline 
\end{tabular}
\par\end{centering}

\caption{Molecular parameters for butadiene. Literature values are taken from
Craig et. al. \cite{Craig2004} (experimental rotational constants)
and Smith et. al. \cite{Smith2004} (calculated static polarizabilities).
Calc. column lists symmetrized values used in the rotational wavepacket
calculations. \label{tab:Molecular-parameters}}
\end{table}

\par\end{center}

\section{Symmetry\label{sub:Symmetry}}

Symmetry plays a role in determining the allowed transition matrix
elements and continuum wavefunctions. Here we treat the case of butadiene
in its equilibrium planar geometry ($C_{2h}$), relevant to recent
time-resolved experiments studying ultra-fast excited state dynamics
from aligned butadiene, which utilised an IR alignment pulse followed
by a 1+1' pump-probe measurement around the peak of the half-revival,
to populate the $S_{2}$ state and probe the vibronic dynamics \citep{hockett2014b}. These
symmetries were used in the limiting case calculations presented
in the main body of this manuscript.

Table \ref{tab:Butadiene-symmetries} lists symmetries for (a) states
of interest, (b) multipoles and (c) dipole transitions. The dipole-allowed
$S_{0}\rightarrow S_{2}$ transition accessed experimentally is $(x,y)$
polarised, i.e. perpendicular to the molecular axis, so corresponds
to a $\sin^{2}\theta$ excitation as detailed in sect. \ref{sub:2-photon-excitation}.
The excited state axis distribution $P'_{\perp}(\theta,t)$ is, therefore,
distinctly different to the originally prepared $P(\theta,t)$. For
1-photon ionization all dipole polarizations are symmetry allowed,
but correspond to different continuum symmetries, hence different
partial waves.

Ultra-fast population transfer from $S_{2}\rightarrow S_{1}$ occurs
in butadiene, and of particular note here is the switch from $g\rightarrow u$
continua for $S_{2}\rightarrow D_{0}$ versus $S_{1}\rightarrow D_{0}$,
corresponding to a switch from even to odd $l$. In this case one
would expect the PADs for these cases to be distinctly different.

As shown in table \ref{tab:Butadiene-symmetries}, the allowed continua
correlate with different $(l,m)$ even/odd combinations. To reduce
the number of matrix elements, linear combinations of $\pm m$ can
be used. In this case the $b_{hl\lambda}^{\Gamma\mu}$ symmetrization
coefficients in eqn. \ref{eq:LF-PAD-t} take values of $1$ for $m=0$,
or $1/\sqrt{2}$ for $m\neq0$. The indices are then reduced to the
set $\Gamma=\{A_{g},\, B_{g}\}$, $\mu=1$ (no degenerate symmetries
present), $h=|m|$ and $\lambda=|m|$.

\begin{table}
\subfloat[State symmetries.]{%
\begin{tabular}{|c|c|}
\hline 
State & Symmetry\tabularnewline
\hline 
\hline 
$S_{0}$ & $^{1}A_{g}$\tabularnewline
\hline 
$S_{1}$ & $^{1}A_{g}$\tabularnewline
\hline 
$S_{2}$ & $^{1}B_{u}$\tabularnewline
\hline 
$D_{0}$ & $^{2}B_{g}$\tabularnewline
\hline 
$D_{1}$ & $^{2}A_{u}$\tabularnewline
\hline 
$D_{2}$ & $^{2}A_{g}$\tabularnewline
\hline 
\end{tabular}

} 

\subfloat[Multipole symmetries in $C_{2h}$.]{%
\begin{tabular}{|c|c|c|}
\hline 
Character & Dipole ($Y_{1m}$) & Multipole ($Y_{lm}$)\tabularnewline
\hline 
\hline 
$A_{g}$ &  & e, e\tabularnewline
\hline 
$B_{g}$ &  & e, o\tabularnewline
\hline 
$A_{u}$ & $z$ & o, e\tabularnewline
\hline 
$B_{u}$ & $(x,\, y)$ & o, o\tabularnewline
\hline 
\end{tabular}

}

\subfloat[Dipole transitions.]{%
\begin{tabular}{|c|c|c|}
\hline 
 & $S_{0}$ & $D_{0}$\tabularnewline
\hline 
\hline 
$S_{0}$ & - & $B_{u}(z)$\tabularnewline
 & - & $A_{u}(x,y)$\tabularnewline
\hline 
$S_{1}$ & - & $B_{u}(z)$\tabularnewline
 & - & $A_{u}(x,y)$\tabularnewline
\hline 
$S_{2}$ & - & $A_{g}(z)$\tabularnewline
 & $(x,y)$ & $B_{g}(x,y)$\tabularnewline
\hline 
\end{tabular}

}

\caption{Butadiene symmetries \& transitions. (a) State symmetries for the first three neutral states $S_{n}$, and first three ionic states $D_{n}$. (b) Multipole symmetries in $C_{2h}$. Characters correlate with different combinations of to $(l,\, m)$, denoted by even (e) or odd (o). Dipole
transition symmetries ($l=1$) are also explicitly given in Cartesian form. (c) Allowed dipole transitions (1 photon) \& polarisation, for bound-bound
transitions $S_{0}\rightarrow S_{2}$, and bound-free transitions
to $D_{0}$ (ionic ground state). Bound-free transitions are labelled
according to the continuum symmetry accessed for different dipole
transition symmetries. As shown in (b), these symmetries correspond
to different sets of partial waves.\label{tab:Butadiene-symmetries}}
\end{table}


\section{Comparison with axis convolution methodology}

Recent work has also considered a similar problem, where alignment
was probed with a strong IR field \cite{Pavicic2007,Mikosch2013,Weber2013}
or via high-harmonic generation \cite{Bertrand2012}. In these cases
the ionization was treated purely geometrically as a (continuous)
convolution of the form:

\begin{equation}
I(\Theta,t)=\iint P(\theta,t)S(\Theta)\sin(\theta)d\theta d\phi
\end{equation}
where $S(\Theta)$ described the angular response of the signal to
the angle between the aligning and probing laser fields, e.g. the
angle-resolved ionization yield, and is given by an expansion in Legendre
polynomials in the usual way: 

\begin{equation}
S(\Theta)=\sum_{L}G_{L}P_{L}(\Theta)
\end{equation}

Conceptually this treatment is very similar to that given here, c.f.
eqns. \ref{eq:BTheta_AKQ_cyl} and \ref{eq:BTheta_AKQ_cyl_YLM}, but
treats the convolution numerically, and incoherently w.r.t. couplings
between $P(\theta,t)$ and $S(\Theta)$, which are both here real
valued functions by definition. For a cylindrically symmetric distribution,
in which all terms in eqn. \ref{eq:BTheta_AKQ_cyl_YLM} are real,
this simplification is valid. More generally (c.f. eqn. \ref{eq:BOmega_AKQ})
this formalism will not correspond to the underlying (molecular frame)
properties of the signal, although it may be possible to find an angular
response function $S(\Theta)$ which still models the observed signal
$I(\Theta,t)$, as this has the form of a generic angular function.
Therefore, an empirical convolution of this form, while mathematically
valid in terms of the symmetry of the problem, should be treated with
care if one is interested in determining physically meaningful properties.

\section{Comparison with CS$_{2}$ formalism}

In previous work on modelling the ionization of CS$_{2}$, specific
equations were derived using a similar geometric treatment of the
ionization yields arising from a 1+1' REMPI process, for both parallel
and perpendicular excitation-probe geometries. The results were \cite{Bisgaard2009}:

\begin{widetext}

\begin{equation}
I_{\parallel}(\theta,\phi)=I_{0}[(r-1)\cos^{4}(\theta)+\cos^{2}(\theta)]
\end{equation}

\begin{equation}
I_{\perp}(\theta,\phi)=I_{0}[\cos^{2}(\theta)+(r-1)\cos^{2}(\theta)\cos^{2}(\phi)-(r-1)\cos^{4}(\theta)\cos^{2}(\phi)]
\end{equation}

\end{widetext}

where $I_{\parallel}(\theta,\phi)$ and $I_{\perp}(\theta,\phi)$
are the angle-resolved ionization yields for a parallel and a perpendicular
excitation-probe polarization geometry respectively, the angles $(\theta,\,\phi)$
refer to the axis alignment relative to the ionization frame, $I_{0}$
is the total yield and $r$ is the ratio of the parallel and perpendicular
ionization dipole moments (similar to the definition used in section
\ref{sub:Limiting-cases}):

\begin{equation}
r=\left(\frac{\mu_{\parallel}}{\mu_{\perp}}\right)^{2}
\end{equation}

These equations show explicitly that $\cos^{4}(\theta)$ terms contribute
in this case. They also indicate the possibility of obtaining $r$
if the alignment distribution is known, or the alignment distribution
if $r$ is known; or, possibly, fitting both in a self-consistent
manner, as was demonstrated in that case based on the measured ratio
of $\langle I_{\parallel}(\theta,\phi)\rangle$ to $\langle I_{\perp}(\theta,\phi)\rangle$
for an unaligned distribution, and the measured time-dependent yield
for ionization from an aligned distribution using a parallel polarization
geometry, $\beta_{0,0}^{\parallel}(t)$ in our notation (eqn. \ref{eq:B00t-2photon-REMPI}).
From the measurement the details of the prepared distribution were
determined by the fitting procedure, where the free parameters were the peak intensity
of the alignment pulse and the rotational temperature used in the
rotational wavepacket calculations - hence the terms $\langle\cos^{2}(\theta,\, t)\rangle$
and $\langle\cos^{4}(\theta,\, t)\rangle$ were found by this fit.
The full axis distribution $P(\theta,\, t)$ based on the best fit
result was later used in modelling of the PADs \cite{Hockett2011}.
Although we have not derived equivalent equations, we note that the
angular terms are of the same order as those appearing in eqns. \ref{eq:B00t-2photon-REMPI}
and \ref{eq:B00t-2photon-REMPI-perp} when expressed in trigonometric
form, so the formalisms may be assumed to be functionally identical.

\bibliographystyle{apsrev4-1}
\bibliography{/media/store/reports/bibliography/hockett_pubs_2013,/media/store/reports/bibliography/alignment_refs_mendeley_030414,/media/store/reports/bibliography/alignment_refs_additional_050513b,/media/store/reports/bibliography/MF-PADs_nature_b,/media/store/reports/bibliography/alignment_refs_additional_2_mendeley_030914}

\end{document}